\documentclass[10pt,a4paper,reqno]{amsart}
\usepackage{acronym}
\usepackage{amsfonts}
\usepackage{amsmath}
\usepackage{amssymb}
\usepackage{amsthm}
\usepackage{bm}
\usepackage{braket}
\usepackage{cite}
\usepackage{color}
\usepackage{geometry}
\usepackage{graphicx}
\usepackage{hyperref}
\usepackage{mathrsfs}
\usepackage{mathtools}
\usepackage{setspace}
\usepackage{stmaryrd}
\usepackage{subcaption}
\usepackage{tikz}
\usepackage{CJKutf8}

\usepackage{epsfig}
\usepackage{setspace}
\usepackage{booktabs}
\usepackage{threeparttable}
\usepackage{diagbox}
\usepackage{float}

\usepackage{booktabs}
\usepackage{threeparttable}
\usepackage{diagbox}
\usepackage{multirow}
\usepackage{float}
\usepackage{array}

\newgeometry{top=2cm,bottom=2cm,outer=1.5cm,inner=1.5cm}
\newcommand{\bse}{\begin{subequations}}
\newcommand{\ese}{\end{subequations}}

\numberwithin{equation}{section}

\title[Data driven solutions and parameter discovery of the nmKdV equation via deep learning method]{Data driven solutions and parameter discovery of the nonlocal mKdV equation via deep learning method}
\author{JinYan Zhu}
\address[JY]{School of Mathematical Sciences, Shanghai Key Laboratory of Pure Mathematics and Mathematical Practice, and Shanghai Key Laboratory of Trustworthy Computing \\
East China Normal University \\ Shanghai 200241 \\ People's Republic of China}
\author{Yong Chen$^*$}
\address[YC]{School of Mathematical Sciences, Shanghai Key Laboratory of Pure Mathematics and Mathematical Practice, and Shanghai Key Laboratory of Trustworthy Computing \\
East China Normal University \\ Shanghai 200241 \\ People's Republic of China}
\address[YC]{College of Mathematics and Systems Science \\ Shandong University of Science and Technology \\ Qingdao 266590 \\ People's Republic of China}
\email{ychen@sei.ecnu.edu.cn}

\begin{document}

\begin{abstract}
In this paper, we systematically study the integrability and data-driven solutions of the nonlocal mKdV equation. The infinite conservation laws of the nonlocal mKdV equation and the corresponding infinite conservation quantities are given through Riccti equation. The data driven solutions of the zero boundary for the nonlocal mKdV equation are studied by using the multi-layer physical information neural network algorithm, which including kink soliton, complex soliton, bright-bright soliton and the interaction between soliton and kink-type. For the data-driven solutions with non-zero boundary, we study kink, dark, anti-dark and rational solution. By means of image simulation, the relevant dynamic behavior and error analysis of these solutions are given. In addition, we discuss the inverse problem of the integrable nonlocal mKdV equation by applying the physics-informed neural network algorithm to discover the parameters of the nonlinear terms of the equation.

\end{abstract}

\maketitle

\section{Introduction}

In recent years, due to the wide application of nonlocal equations in various aspects, it has attracted the attention of many scholars\cite{SN-2012-PRA,AMJ-2013-PRL,CM-2002-PRL}. The most classical nonlocal equation was the nonlocal nonlinear Schr\"{o}dinger(NNLS) equation proposed by Ablowitz and Musslimani in 2013\cite{AMJ-2013-PRL}. And the most important parity-time(PT) symmetry in nonlocal equations was introduced into AKNS system for the first time. Later, many scholars have studied this equation at various levels\cite{AMJ-2016-N,FBF-2018-N,WM-2021-ND,RY-2019-JMP,LGZ-2021-AR}. Then many nonlocal equations are proposed and studied, such as nonlocal Davey-Stewartson equation\cite{JGR-2017-SAM}, nonlocal derivative nonlinear Schr\"{o}dinger equation\cite{ZZX-2018-CNSNS}, nonlocal Hirota\cite{CJ-2019-JMP,ZJ-2022-AR}, nonlocal KdV \cite{GM-2020-PLA} and so on. Recently, a nonlocal modified KdV(mKdV) was proposed
\begin{equation}\label{mkdv}
u_{xxx}+u_{t}+6uu(-x,-t)u_{x}=0,
\end{equation}
which also called reverse-space-time mKdV equation. In physical applications, the nonlocal mKdV has shifted parity and delayed time reversal symmetry, which is related to Alice Bob system\cite{SYL-2017-SR}. In fact, the nonlocal mKdV equation has been widely discussed by scholars. For example, the inverse scattering transform of the nonlocal mKdV equation was given in\cite{JLJ-2017-JMAA,ZG-2020-PD}. The soliton solution of the nonlocal mKdV equation was solved through Darboux transformation \cite{JLJ-2017-CNSNS}. The Dbar dressing method for the nonlocal mKdV equation was shown in Ref.\cite{LJ-2022-JGP}. The long-time asymptotic behavior of the nonlocal mKdV equation with decaying initial data was studied by using Deift-Zhou steepest descent method\cite{FJH-2022-JGP}.

Conservation law is universal in applied mathematics\cite{HPE-2001-JPA}. It reflects a phenomenon that some physical quantities do not change with time. In soliton theory, conservation law plays an important role in discussing the integrability of soliton equations. The existence of infinitely many conservation laws is closely related to the existence of soliton equations. In fact, most nonlinear development equations with soliton solutions have infinitely many conservation laws. Therefore, for a soliton system, finding its infinite conservation law is of great practical and theoretical significance for proving the integrability of the system. Since Miura, Gardner and Kruskal discovered that the KdV equation has an infinite conservation law\cite{MRM-1968-JMP}, a series of methods have been developed to construct the (1 + 1)-dimensional integrable system, some of which are no longer in use due to their limitations. For instance, through the scattering problem and the gradual expansion of the scattering quantity $a(\lambda)$ can yield an endless number of conserved quantities\cite{AMJ-1976-JMP}, but it cannot be used to build the conservation rule, hence this method is currently of little use in application areas. In the study of infinite conservation law, Wadati et al. have made considerable contributions. Generally speaking, the infinite conservation law of a continuous system can be obtained through the following ways: Lax pair, B\"{a}cklund transform, formal solution of eigenfunction and trace identity, etc\cite{WM-1975-P,SA-1972-SPJ,KK-1974-PTP,TT-1998-JPS}. Although there are many ways to obtain infinite conservation law, the conservation law is the same. This variety of methods and the consistency of results can be seen as an external manifestation of integrability.

Machine learning is the mainstream method to solve many AI problems at this stage. As an independent direction, it is developing at a high speed. As a form of machine learning, deep learning trains models with multiple hidden layers between input and output. In general, deep learning means using deep neural networks. Deep neural network has the advantages of fast computing speed and high accuracy, and has been widely used in natural language processing\cite{CR-2011-JM}, face recognition\cite{SXD-2018-N}, speech recognition\cite{NAB-2019-IEEE} and other fields\cite{AB-2015-NB,SZF-2020-IEEE}. Recently, A new physical information neural network (PINN) is proposed by the mathematical physics system based on the multi-layer network of deep learning mode, which is proved to be suitable for dealing with forward problems and highly ill inverse problems. The approximate solution of the control equation and the parameters of the control equation are found from the training data\cite{MR-2019-JCP}. And the numerical findings demonstrate that high-dimensional network tasks can be successfully completed using the PINN approach with fewer data sets. This training neural network is a supervised learning task to solve some nonlinear partial differential equations that follow the laws of physics.
Then, the PINN method is used to generate data-driven solutions to reveal the dynamic behavior of nonlinear partial differential equations under physical constraints, which has attracted wide attention from many scholars. Chen's team has built many data-driven solutions of nonlinear partial differential equations using the PINN approach during the last two years, including soliton solutions\cite{JL-2020-CTP,PWQ-2022-ND}, breather solutions\cite{JCP-2021-CPB}, rational solution\cite{JCP-2021-ND},  rogue wave\cite{JCP-2022-CSF,JCP-2021-CPB}, higher-order breather waves\cite{ZWM-2021-MPLB}and rogue periodic wave\cite{WQP-2022-CNSNS}. In particular, Lin and Chen add the properties of integrable systems such as conserved quantities and Miura transform to the training network and proposes a two-stage PINN method and finds new local wave solutions\cite{LS-2022-JCP,LS-2022-AR}. In addition, other scholars have also used PINN method to obtain some important results of data-driven solutions concerning the defocusing NLS equation with potential energy and the coupled NLS equation\cite{WL-2021-PLA,FY-2021-ND,MY-2022-PLA}. As far as we know, the application of PINN to nonlocal equations has been rarely studied.


In this paper, we will derive the Ricatti equation from the $x$ part of the Lax pair of the nonlocal mKdV equation and construct the conservation law using the compatibility condition. The infinite conservation laws and infinite conserved quantities are obtained from the solution of Ricatti equation. Besides, we add the nonlocal term to the classical PINN to simulate the data-driven solutions of the nonlocal mKdV equation under zero boundary and nonzero boundary conditions and give the error analysis. At the same time, we use the PINN of the nonlocal term to discover the parameters of the nonlocal mKdV.

The structure of this paper is as follows. In Sec. 2, we mainly study the integrability of the nonlocal mKdV equation and obtain its infinite conservation laws by using the Riccti equation. In Sec. 3, the composition of PINN is introduced. Then we use PINN to study the data-driven solution under zero boundary conditions and give its dynamic behavior in Sec. 4. In Sec. 5, we learn the data-driven solutions of the nonzero boundary of the nonlocal mKdV equation and its dynamic behavior. In Sec. 6, the inverse problem is learned based on PINN, which mainly studies the nonlinear coefficients of the nonlocal mKdV equation, while adding different noises to the network. The conclusion is given in Sec.7.

\section{Conservation laws for the nonlocal mKdV equation}
The nonlocal mKdV equation (\ref{mkdv}) has the following Lax pair
\begin{equation}\label{lax}
\begin{aligned}
&\Phi_x=M \Phi, \quad M=M(x, t ; \lambda):=i \lambda \sigma_3+U, \\
&\Phi_t=N \Phi, \quad N=N(x, t ; \lambda):=\left[4 \lambda^2-2 u(x, t) u(-x,-t)\right] M-2 i \lambda \sigma_3 U_x+\left[U_x, U\right]-U_{x x},
\end{aligned}
\end{equation}
where $\Phi$ is the matrix eigenfunction, $\lambda$ is the spectral parameter and
$$
\sigma_3=\left[\begin{array}{cc}
1 & 0 \\
0 & -1
\end{array}\right],~~U(x, t)=\left[\begin{array}{cc}
0 & u(x, t) \\
-u(-x,-t) & 0
\end{array}\right].
$$
For the solution of Lax pair equation (\ref{lax}) written in the form of vector $\Phi=(\phi_1,\phi_2)^T$, the following function is introduced
$$
\Gamma=\frac{\phi_2}{\phi_1},
$$
we can put it into equation (\ref{lax}) and get
$$
\begin{aligned}
&(ln{\phi_1})_x=i\lambda+u\Gamma,\\
&(ln{\phi_1})_t=4i\lambda^3-2iuu(-x,-t)\lambda-u_xu(-x,-t)-u_x(-x,-t)u+(4\lambda^2u-2u^2u(-x,-t)-2i\lambda u_x-u_{xx})\Gamma.
\end{aligned}
$$
The derivatives of $x$ and $t$ of the above two equations are as follows
\begin{equation}\label{cl}
(u\Gamma)_t=(-2iuu(-x,-t)\lambda-u_xu(-x,-t)-u_x(-x,-t)u+(4\lambda^2u-2u^2u(-x,-t)-2i\lambda u_x-u_{xx})\Gamma)_x.
\end{equation}
In addition, combining the Lax pair equation, we can get the Riccti equation and conservation law equation about $\Gamma$
$$
\begin{aligned}
&\Gamma_x=-u(-x,-t)-2i\lambda\Gamma-u\Gamma^2,\\
&\Gamma_t=-4\lambda^2u(-x,-t)+2uu^2(-x,-t)+2i\lambda u_x(-x,-t)+u_{xx}(-x,-t)+A\Gamma-B\Gamma^2,
\end{aligned}
$$
where
$$
A=-8i\lambda^3+4i\lambda uu(-x,-t)+2u_xu(-x,-t)+2uu_x(-x,-t),~~B=4\lambda^2u-2u^2u(-x,-t)-2i\lambda u_x-u_{xx}.
$$
The function $\Gamma$ is expanded in the following series
$$
\Gamma(x, t, \lambda)=\sum_{n=1}^{\infty} \frac{\Gamma^{(n)}(x, t, \lambda)}{(2 i \lambda)^n},
$$
and collect the coefficient of the same power of $\lambda$ to obtain the following formula
$$
\begin{aligned}
&\Gamma^{(1)}=-u(-x,-t),~~~\Gamma^{(2)}=-u_x(-x,-t),\\
&\Gamma^{(3)}=-u_{xx}(-x,-t)-uu^2(-x,-t),\\
&\Gamma^{(4)}=u_xu^2(-x,-t)-4uu(-x,-t)u_x(-x,-t)-u_{xxx}(-x,-t),\\
&\Gamma^{(2n)}=-\Gamma^{(2n-1)}_x-2u\sum_{l+k=2n-1}\Gamma^{(l)}\Gamma^{(k)},\\
&\Gamma^{(2n+1)}=-\Gamma^{(2n)}_x-u\left({\Gamma^{(n)}}^2+2\sum_{l+k=2n}\Gamma^{(l)}\Gamma^{(k)}\right).
\end{aligned}
$$
Thus, we can write several conservation laws for the nonlocal mKdV equation as
$$
\begin{aligned}
&(-uu(-x,-t))_t=(3u^2u^2(-x,-t)+u_{xx}u(-x,-t)+uu_{xx}(-x,-t)+u_xu_x(-x,-t))_x,\\
&(uu_x(-x,-t))_t=(-6u^2u_x(-x,-t)u(-x,-t)-u_xu_{xx}(-x,-t)-u_{xx}u_x(-x,-t)-uu_{xxx}(-x,-t))_x,\\
&(-uu_{xx}(-x,-t)-u^2u^2(-x,-t))_t=(-4u^2u(-x,-t)u_{xx}(-x,-t)+4u^3u^3(-x,-t)+u_{xx}u_{xx}(-x,-t)-u^2_xu^2(-x,-t)\\
&~~~~~~~~~~~~~~~~~~~~~~~-2uu_xu(-x,-t)u_x(-x,-t)+uu_{xx}u^2(-x,-t)-2uu_xu(-x,-t)u_x(-x,-t)+u_xu_{xxx}(-x,-t)\\
&~~~~~~~~~~~~~~~~~~~~~~~+5u^2u_x^2(-x,-t)+uu_{4x}(-x,-t))_x,\\
&~~~~~.~~~.~~~.~~~,\\
&(u\Gamma^{(n)})_t=((-2u^2u(-x,-t)-u_{xx})\Gamma^{(n)}-u_{x}\Gamma^{(n+1)}-u\Gamma^{(n+2)})_x.
\end{aligned}
$$
At the same time, the corresponding conserved quantity can also be obtained as
\begin{equation}\label{i1}
\begin{aligned}
&I_1=\int_{-\infty}^{+\infty}-uu(-x,-t)dx,\\
&I_2=\int_{-\infty}^{+\infty}-uu_x(-x,-t)dx,\\
&I_3=\int_{-\infty}^{+\infty}(uu_{xx}(-x,-t)-u^2u^2(-x,-t))dx,\\
&~~~~~.~~.~~.~~,\\
&I_n=\int_{-\infty}^{+\infty}u\Gamma^{(n)}dt.
\end{aligned}
\end{equation}
The above proves the integrability of the nonlocal mKdV equation. Next, we will apply PINN to the integrable nonlinear equation to obtain its data-driven solutions.

\section{The PINN deep learning method}\label{Methodology}
In this part, we will introduce the PINN deep learning method for partial differential equation. Generally, the (1+1)-dimensional nonlinear partial differential equation has the following form
\begin{align}\label{ut}
u_{t}+\mathcal{N}[{u}]=0,~x\in [x_0,x_1],~t\in [t_0,t_1]
\end{align}
where $u$ is a function of $x$ and $t$, $\mathcal{N}[\cdot]$ is a nonlinear differential operator in space, which generally contains high-order dispersion terms and nonlinear terms. Then, the following equation can be defined by the left part of Eq. (\ref{ut})
\begin{equation}\label{f}
f:=u_{t}+\mathcal{N}[{u}].
\end{equation}
and a deep neural network is used to approximate $u$. Here, without losing generality, we first consider a neural network with depth $H$, which is composed of an input layer, $H-1$ hidden layers and an output layer. And the $h^{th}$ hidden layer is composed of $N_h$ neurons, and then the output $\mathbf{x}^{h-1}$ of the previous layer after the action of the activation function $\sigma$ is taken as the input of the next hidden layer. This process is formed through the following radiation transformation
$$
\mathbf{x}^{h}=\sigma\left({\Lambda}_h\left(\mathbf{x}^{h-1}\right)\right)=\sigma\left(\mathbf{w}^h \mathbf{x}^{h-1}+\mathbf{b}^h\right),
$$
where $\mathbf{w}^h\in\mathbb{R}^{N_h\times N_{h-1}}$  and $\mathbf{b}^h\in \mathbb{R}^{N_h}$ are the weights and deviations of the $h^{th}$ layer. Usually, we initialize the bias term to zero, the weight is initialized by Xavier initialization, and the activation function is selected as $tanh$ function. After layer upon layer operation, a neural network can be obtained as
$$
u\left(\mathbf{x}^0, \Theta\right)=\left(\Lambda_H \circ \sigma \circ \Lambda_{H-1} \circ \cdots \circ \sigma \circ \Lambda_1\right)\left(\mathbf{x}^0\right)
$$
where the operator $"\circ"$ is the composition operator, and $\Theta=\left\{\mathbf{w}^h, \mathbf{b}^h\right\}_{h=1}^H$ represents the parameters that can be learned in the network. The core of the neural network is to constantly update the weights and deviations so that the solution $u$ of the partial differential equation satisfies Eq. (\ref{f}) and minimizes $f$. And the neural network $f$ and the network representing $u$ have the same parameters, and these shared parameters can be learned by minimizing the mean square error loss
\begin{equation}\label{l1}
\operatorname{Loss}_{1}=\operatorname{Loss}_u+\operatorname{Loss}_{f},
\end{equation}
where
\begin{align}\label{lu}
&Loss_{u}=\frac{1}{N_{u}}\sum_{i=1}^{N_{u}}|u(x_{u}^{i},t_{u}^{i})-u^{i}|^{2},\\
&Loss_{f}=\frac{1}{N_{f}}\sum_{j=1}^{N_{f}}|f(x_{f}^{j},t_{f}^{j})|^{2},
\end{align}
where $\{x_{u}^{i}, t_{u}^{i}, u^{i}\}_{i=1}^{N_{u}}$ is the sampled initial and boundary value  training data of $u(x, t)$. Similarly, the collocation points for $f(x, t)$ is marked by $\{x_{f}^{j}, t_{f}^{j}\}_{j=1}^{N_{f}}$. The loss function \eqref{l1} contains the losses of
initial-boundary value data and the losses of  networks \eqref{ut} at a finite set of collocation points.

Further, if the solution of the partial differential equation is complex, we can write the solution of the final form as $u=p+iq$.
Thus, the Eq. \eqref{l1} can be divided into two equations with real part and imaginary part, as follows
\begin{align}\label{pt}
p_{t}+\mathcal{N}[p]=0,
\end{align}
\begin{align}\label{qt}
q_{t}+\mathcal{N}[q]=0.
\end{align}

Then the physics-informed neural networks $f_{p}(x, t)$ and $f_{q}(x, t)$ can be defined as
\begin{align}\label{fp}
f_{p}:=p_{t}+\mathcal{N}[p],
\end{align}
\begin{align}\label{fq}
f_{q}:=q_{t}+\mathcal{N}[q],
\end{align}
where $p(x, t; w, b)$, $q(x, t; w, b)$ are the latent function of the deep neural network with the weight parameter $w$ and bias parameter $b$, which can be used to approximate the exact complex-valued solution $u(x, t)$ of objective equations. This form is reasonable, and in Ref.\cite{AG-2018-JMLR} the network $f_{p}(x, t), f_{q}(x, t)$ can be found under the automatic differentiation mechanism.

Similarly, there is also a loss function in this network, and its form is more complex. We can set it to the following form
\begin{align}\label{l2}
Loss^{'}_{1}=Loss_{p}+Loss_{q}+Loss_{f_{p}}+Loss_{f_{q}},
\end{align}
where
\begin{align}\label{lpq}
Loss_{p}=\frac{1}{N_{p}}\sum_{i=1}^{N_{p}}|p(x_{p}^{i},t_{p}^{i})-p^{i}|^{2},\quad Loss_{q}=\frac{1}{N_{q}}\sum_{i=1}^{N_{q}}|q(x_{q}^{i},t_{q}^{i})-q^{i}|^{2},
\end{align}
and
\begin{align}\label{fpq}
Loss_{f_{p}}=\frac{1}{N_{f}}\sum_{j=1}^{N_{f}}|f_{p}(x_{f}^{j},t_{f}^{j})|^{2},\quad
Loss_{f_{q}}=\frac{1}{N_{f}}\sum_{j=1}^{N_{f}}|f_{q}(x_{f}^{j},t_{f}^{j})|^{2},
\end{align}
where $\{x_{p}^{i}, t_{p}^{i}, p^{i}\}_{i=1}^{N_{p}}$ and $\{x_{q}^{i}, t_{q}^{i}, q^{i}\}_{i=1}^{N_{q}}$ are the sampled initial and boundary value  training data of $u(x, t)$. Similarly, the collocation
points for $f_{u}(x, t)$ and $f_{v}(x, t)$ are marked by $\{x_{f}^{j}, t_{f}^{j}\}_{j=1}^{N_{f}}$ and $\{x_{f}^{j}, t_{f}^{j}\}_{j=1}^{N_{f}}$.

For the loss function (\ref{l2}), the first two terms try to make the learned solution close to the exact solution when approaching the initial value and boundary value data, and the last two terms make the hidden functions $p$ and $q$ meet Eqs. (\ref{fp}) and (\ref{fq}).

\section{Soliton solutions of nonlocal mKdV equation under the condition of zero boundary}
\quad

In this part, we mainly use the neural network method to obtain the simulation solution of the nonlocal mKdV equation under the zero boundary condition, as well as its dynamic behavior and error analysis. We consider the nonlocal mKdV equation with Dirichlet boundary conditions, which is given by the following formula:
\begin{align}\label{mkdvu}
\left\{
\begin{array}{l}
u_{xxx}+u_{t}+6uu(-x,-t)u_{x}=0, \quad x\in [x_{0}, x_{1}], t\in [t_{0}, t_{1}],\\
u(x, t_{0})=u_{0}(x),\\
u(x_{0}, t)=u_{1}(t),\ u(x_{1},t)=u_{2}(t),
\end{array}
\right.
\end{align}
where $x_{0}, x_{1}$ represent the boundary of $x$, and $t_{0}, t_{1}$ represent the start and end times of time $t$. The $u_{0}(x)$ defines the initial condition.

\begin{figure}[htbp]
\centering
\includegraphics[width=14cm,height=8cm]{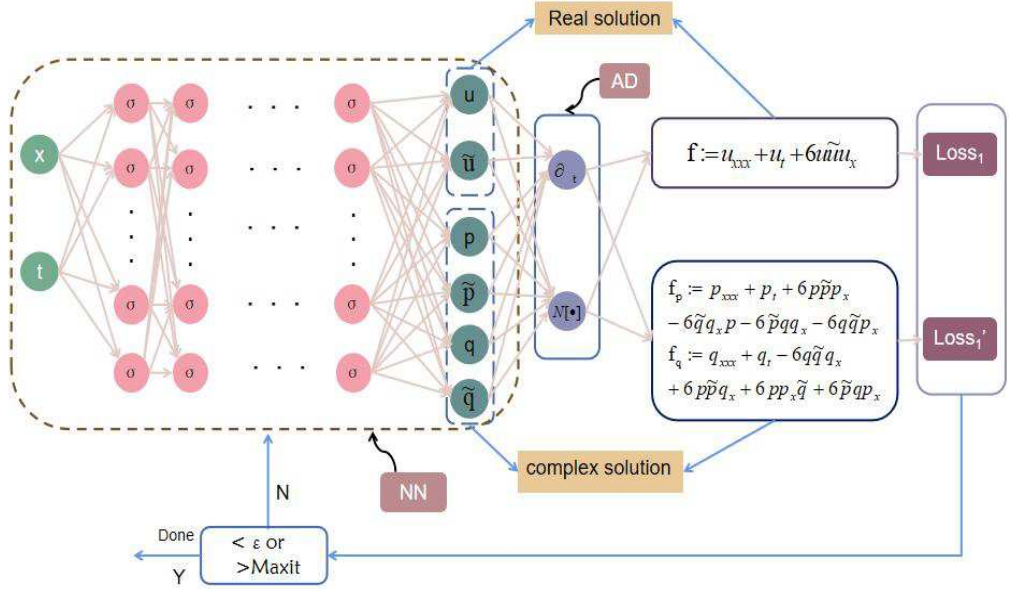}\\
\caption{(Color online)  The PINN scheme solving the nonlocal mKdV equation, where $\tilde{u}=u(-x,-t), \tilde{p}=p(-x,-t)$ and $\tilde{q}=q(-x,-t)$. }\label{lct}
\end{figure}

In order to better understand the PINN method, we give the flow diagram of the nonlocal mKdV equation according to section 3. It can be seen from Fig.\ref{lct} that compared with the classical network diagram of the local equation, the nonlocal term is added to the NN part, which makes more functions output and more complex training. Then the relevant physical information is supplemented, and the loss function is evaluated by NN and the residual of the control equation given in combination with the relevant physical information. Then, the weight $w$ and the deviation $b$ are continuously updated to minimize the loss function to less than a certain tolerance $\varepsilon$ until the specified maximum number of iterations is reached.

Next, we will use physical neural networks to simulate the 1-soliton and 2-soliton solutions of the nonlocal mkdV equation, including kink solution, complex solution and their interactions. At the same time, the simulated solution is compared with the accurate solution.

\subsection{1-soliton solution}
In Ref.\cite{JLJ-2017-CNSNS}, the method of Darboux transformation is used to obtain the kink solution form of the nonlocal mKdV:
$$
u(x, t)=\frac{-2 \nu}{1+e^{-2 \nu\left(x-4 \nu^2 t\right)}}
$$
when $\nu=-1$, a kink solution is generated,
\begin{equation}
u(x,t)=\frac{2}{1+e^{2x-8t}},
\end{equation}
which is a real solution. Therefore, according to Eq. (\ref{f}), the PINN $f(x, t)$ can be constructed as
\begin{equation}\label{fu}
f:=u_{xxx}+u_{t}+6uu(-x,-t)u_{x},
\end{equation}
here we choose [-4,4] as the boundary of $x$ and $t$, so we can give the initial and boundary information as follows:
\begin{align}\label{bu}
\left\{
\begin{array}{l}
u_0(x)=u(x,-4)=\frac{2}{1+e^{2x+32}},\\
~\\
u_1(t)=u(-4,t)=\frac{2}{1+e^{-8-8t}},\\
~\\
u_2(t)=u(4,t)=\frac{2}{1+e^{8-8t}}.
\end{array}
\right.
\end{align}
To get a better simulation effect, we divide $x$ into 512 points in the interval and 200 points in the time interval. That is to say, the interval is divided into $512\times200$ points. $N_u=100$ points are randomly selected in the initial boundary data set, and the internal $N_f=10000$ points are sampled by the Latin hypercube sampling method \cite{SM-1987-T}. Here we construct a feedforward neural network with 6 layers and 40 neurons in each hidden layer. By adjusting all the learnable parameters of the neural network and the loss function, we successfully learn the 1-soliton solution $u(x, t)$. The relative $\mathbb{L}_2$ error of the final PINN model is $2.583821e-04$ in about 84.7319s, and the number of iterations is 346.

\begin{figure}[htbp]
\centering
\includegraphics[width=12cm,height=3.5cm]{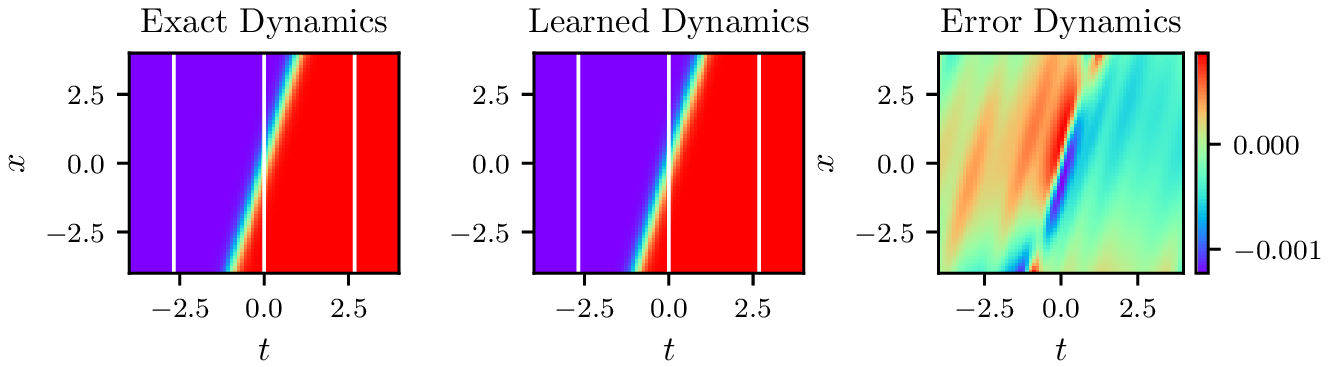}\\
 \quad \quad \quad \quad \quad \quad  $(a)$\\
\includegraphics[width=12cm,height=4.5cm]{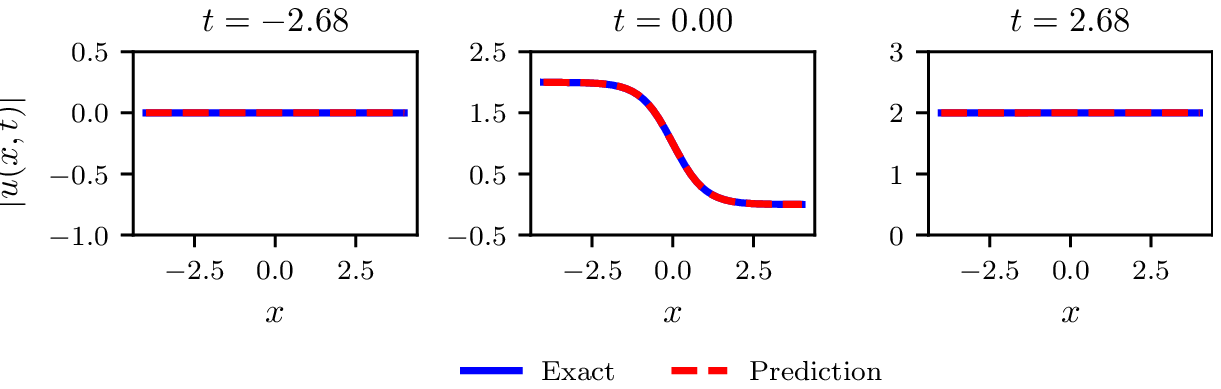}\\
\quad \quad \quad \quad \quad \quad $(b)$\\
\includegraphics[width=7cm,height=5cm]{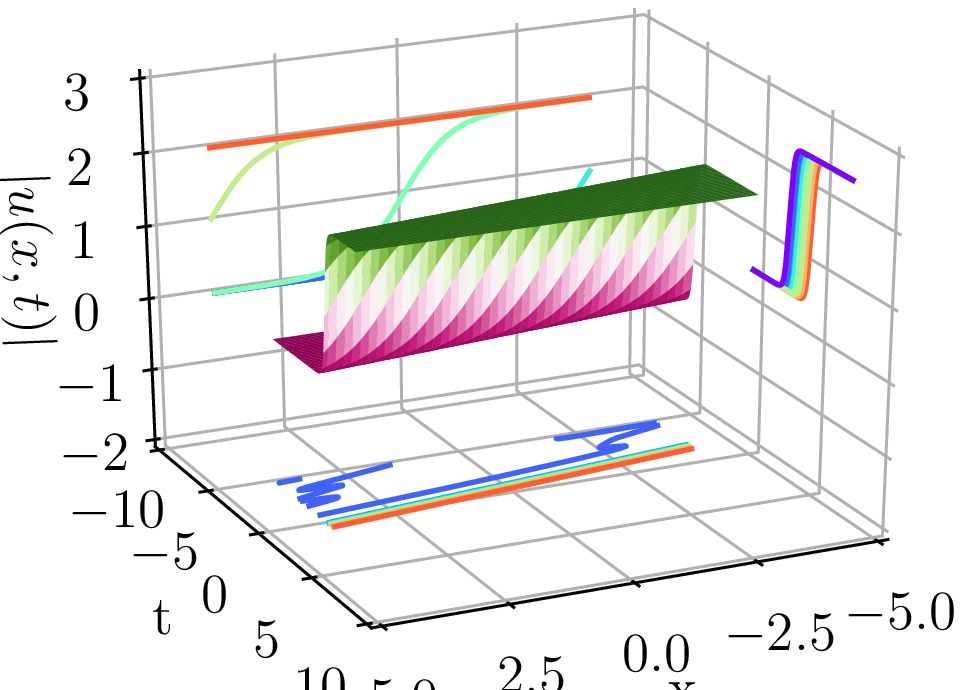}
\includegraphics[width=7cm,height=5cm]{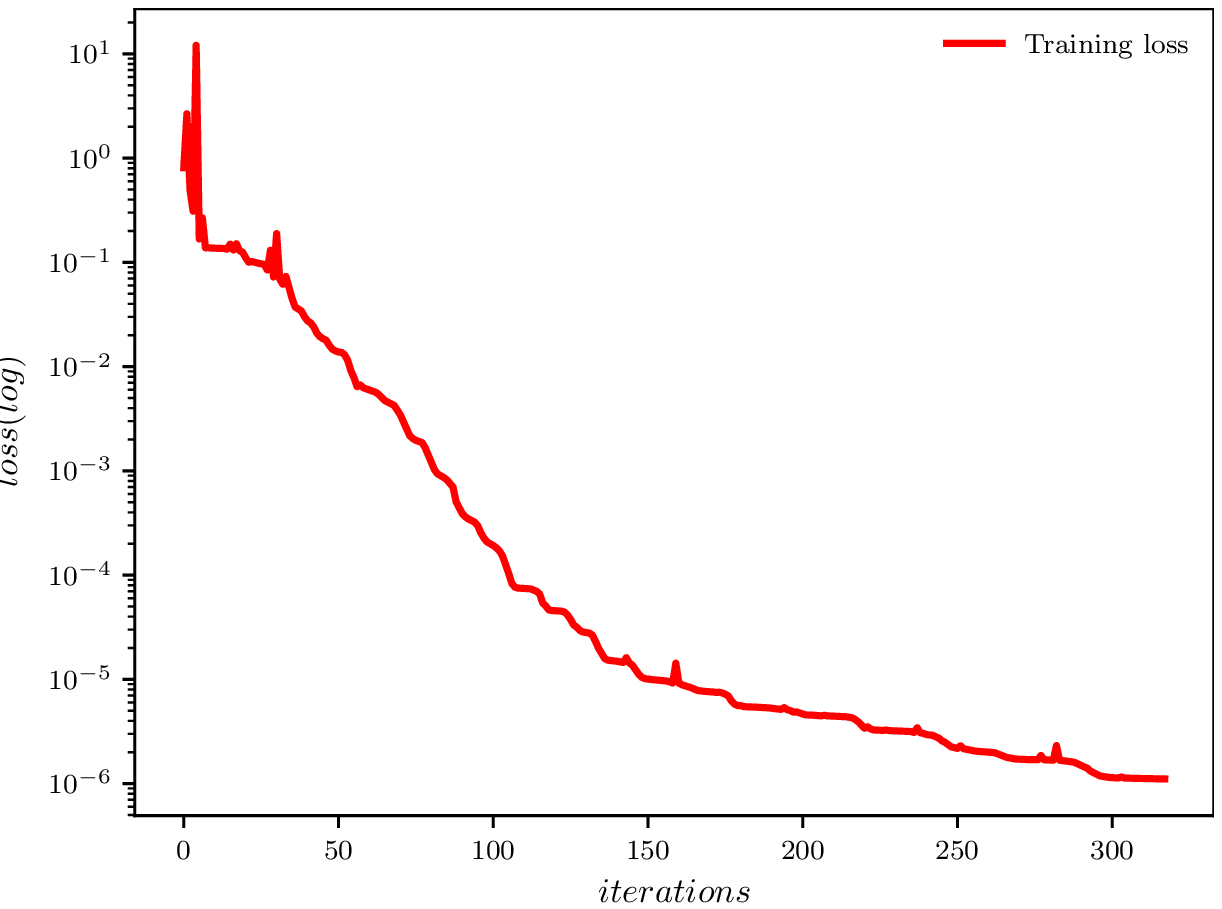}\\
$(c)$ \quad \quad \quad \quad \quad \quad \quad \quad \quad \quad $(d)$
\caption{(Color online) The 1-soliton solution $u(x, t)$ for the nonlocal mKdV equation:(a) The density  plot and the  error density diagram;
(b) The wave propagation plot at three different times;
(c) The three-dimensional plot;
(d) The loss curve figure.}\label{1t}
\end{figure}
Fig.\ref{1t} (a) and (b) show the density plots of the exact solution and the learning solution, the error density plots and three wave propagation plots at different times respectively. We can find that the error between the learning solution and the accurate solution is very small. Fig. \ref{1t} (c) and (d) display the three-dimensional motion and loss curves respectively. As shown in Fig.\ref{1t} (d), the loss curve is relatively smooth, which implies that the integrable deep learning method is very effective and stable.

\begin{figure}[htbp]
\centering
\includegraphics[width=12cm,height=3.5cm]{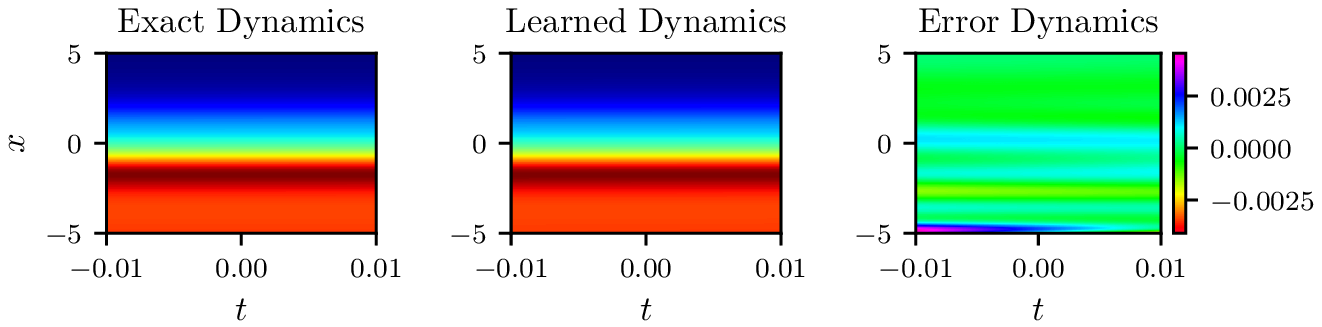}\\
\quad \quad \quad \quad   $(a)$\\
\includegraphics[width=12cm,height=4.5cm]{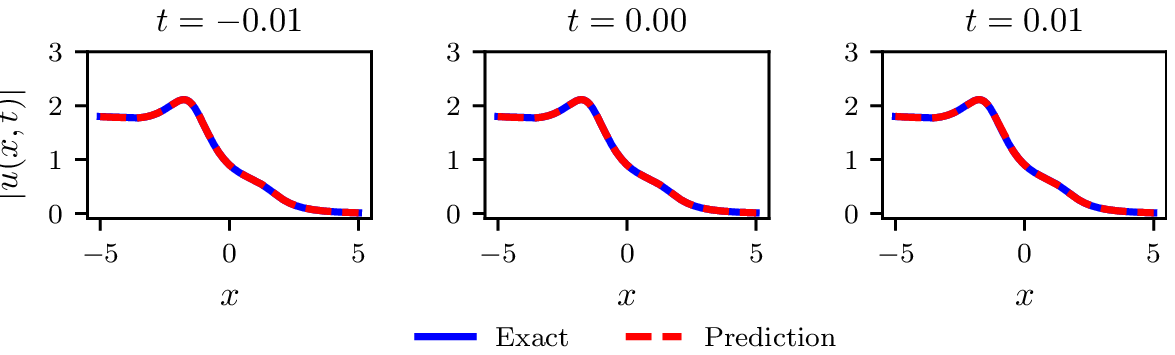}\\
\quad \quad \quad \quad  $(b)$\\
\includegraphics[width=7cm,height=5cm]{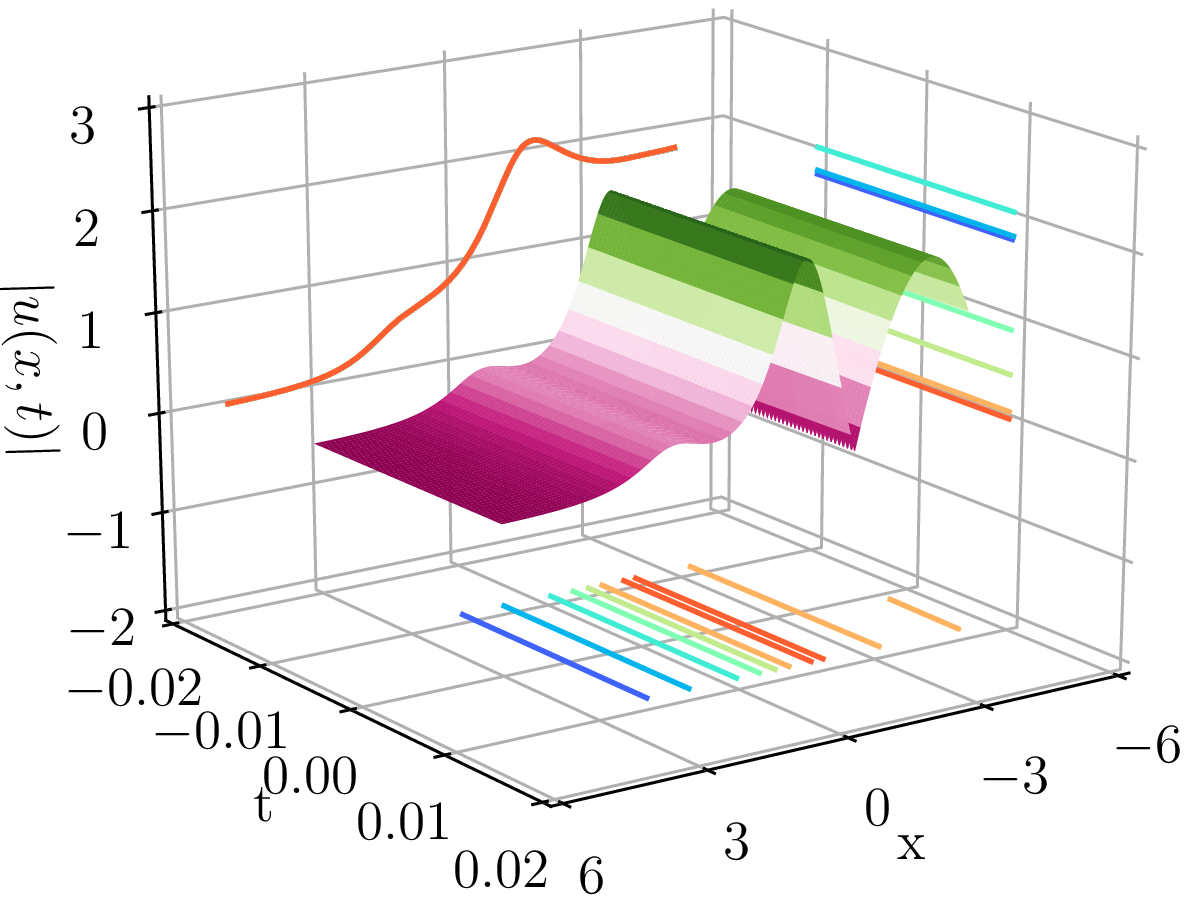}
\includegraphics[width=7cm,height=5cm]{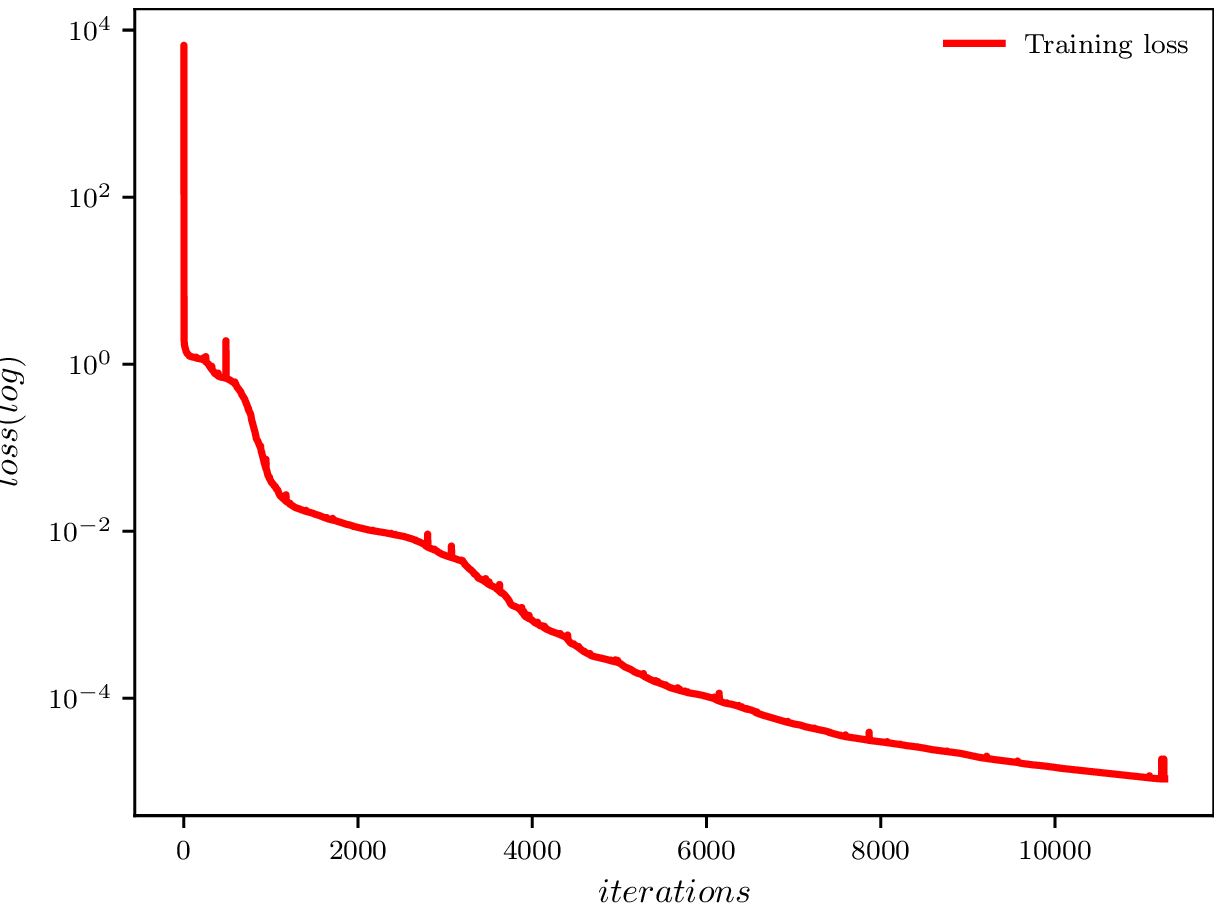}\\
$(c)$ \quad \quad \quad \quad \quad \quad \quad \quad \quad \quad \quad $(d)$
\caption{(Color online) The complex soliton solution $u(x, t)$ for the nonlocal mKdV equation:(a) The density  plot and the  error density diagram;
(b) The wave propagation plot at three different times;
(c) The three-dimensional plot;
(d) The loss curve figure.}\label{2t}
\end{figure}

In addition,  the complex soliton solution of the nonlocal mKdV equation was given by the Hirota direct method in Ref. \cite{GM-2019-CNSNS}, which can be expressed as
\begin{equation}\label{u1f}
u(x,t)=\frac{(1+\frac{3}{2}i)e^{ix+it}}{1+e^{(1+\frac{3}{2}i)x-(\frac{1}{4}+\frac{3}{8}i)t}}.
\end{equation}
In this case, we set $u=p+iq$, then the construction of PINN needs to divide the real part and the imaginary part of the equation reference Eqs.(\ref{fp}) and (\ref{fq}),
\begin{align}\label{fpf}
f_{p}:=p_{xxx}+p_{t}+6pp(-x,-t)p_{x}-6q(-x,-t)q_xp-6p(-x,-t)qq_x-6qq(-x,-t)p_x,
\end{align}
\begin{align}\label{fqf}
f_{q}:=q_{xxx}+q_{t}-6qq(-x,-t)q_{x}+6pp(-x,-t)q_x+6pp_xq(-x,-t)+6p(-x,-t)qp_x.
\end{align}
Let $[x_{0}, x_{1}]$ and $[t_{0}, t_{1}]$ in Eq.\eqref{mkdvu} as $[-5, 5]$ and $[-\frac{1}{100}, \frac{1}{100}]$ respectively. We select the complex soliton solution at $t=-\frac{1}{100}$ as the initial condition
\begin{align}\label{u0f}
u_0(x)=u(x,-\frac{1}{100})=\frac{(1+\frac{3}{2}i)e^{ix-\frac{1}{100}i}}{1+e^{(1+\frac{3}{2}i)x+\frac{1}{400}+\frac{3}{800}i}}.
\end{align}
With the help of LHS, $N_u=100$ collocation points and $N_f=5000$ collocation points are randomly selected at the boundary and inside, respectively, to obtain training data, and put them into a network with 6 hidden layers and 40 neurons. Finally, we successfully learned the complex soliton solution, and the training solution and accurate solution achieved a $\mathbb{L}_{2}$ error of 5.639394e-04. The whole training process take 743.5638 seconds, with 10414 iterations. In Fig. \ref{2t}, we not only give the density and error diagrams of the exact solution and the training solution, but also show the differences between the two solutions at different time stages. It can be seen from Fig.\ref{2t} (b) that they are very consistent. In addition, we also give the 3-dimensional graph of the training solution and the loss function graph in the iteration process in graphs (c) and (d).

\subsection{2-soliton solution}

In this subsection, we mainly simulate two soliton solutions, including bright-bright soliton and soliton kink interaction solutions. As shown in Ref.\cite{GM-2019-CNSNS}, the general formula of the 2-soliton solution is
\begin{equation}\label{u20}
u(x,t)=\frac{F}{G},
\end{equation}
where
$$
\begin{aligned}
F=&\frac{a_1a_2\left(\frac{3}{2}(k+\frac{1}{2}) b_1 e^{(1+k+l) x-(k^3+l^3+1) t}+(1+l)(k+l) b_2 e^{\left(\frac{3}{2}+k\right) x-\left(k^3+\frac{9}{8}\right) t}\right)}{l-\frac{1}{2}}+\frac{\frac{3}{2} a_1(1+l)e^{-t+x}+a_2(k+\frac{1}{2})(k+l) e^{-k^3 t+k x}}{1-k},\\
G=&1+\frac{1}{(1-k)\left(l-\frac{1}{2}\right)}\left(\frac{3}{2} a_1 b_1(k+l) e^{(1+l) x-\left(l^3+1\right) t}+a_1 b_2(1+l)(k+\frac{1}{2}) e^{\frac{3}{2} x-\frac{9}{8} t}+a_2 b_1(k+\frac{1}{2})(1+l) e^{(k+l) x-\left(k^3+l^3\right) t}\right. \\
&\left.+\frac{3}{2} a_2 b_2(k+l) e^{\left(k+\frac{1}{2}\right) x-\left(k^3+\frac{1}{8}\right) t}\right)+a_1 a_2 b_1 b_2 e^{\left(\frac{3}{2}+k+l\right) x-\left(k^3+l^3+\frac{9}{8}\right) t}.
\end{aligned}
$$
When $k=\frac{1}{2},l=1,a_1=a_2=b_1=b_2=1$, Eq.(\ref{u20}) can be reduced to
\begin{equation}
u(x,t)=\frac{6 e^{x-t}+3 e^{\frac{1}{2} x-\frac{1}{8} t}+3 e^{\frac{5}{2} x-\frac{17}{8} t}+6 e^{2 x-\frac{5}{4} t}}{1+9 e^{2 x-2 t}+16 e^{\frac{3}{2} x-\frac{9}{8} t}+9 e^{x-\frac{1}{4} t}+e^{3 x-\frac{9}{4} t}},
\end{equation}
which is a non-singular 2-soliton solution. Let $[x_{0}, x_{1}]$ and $[t_{0}, t_{1}]$ in Eq.\eqref{mkdvu} as $[-15, 15]$,  the corresponding initial condition is given by
\begin{align}
u_{0}(x)=u(x,-15)=\frac{6 e^{x+15}+3 e^{\frac{1}{2} x+\frac{15}{8}}+3 e^{\frac{5}{2} x+\frac{255}{8}}+6 e^{2 x+\frac{75}{4}}}{1+9 e^{2 x+30}+16 e^{\frac{3}{2} x+\frac{135}{8}}+9 e^{x+\frac{15}{4}}+e^{3 x+\frac{135}{4}}}.
\end{align}

\begin{figure}[htbp]
\centering
\includegraphics[width=12cm,height=3.5cm]{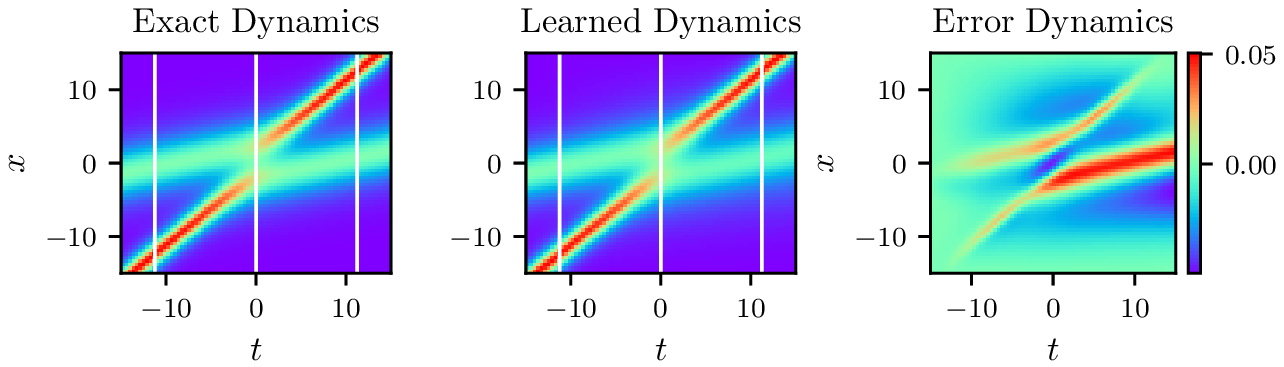}\\
\quad \quad \quad \quad   $(a)$\\
\includegraphics[width=12cm,height=4.5cm]{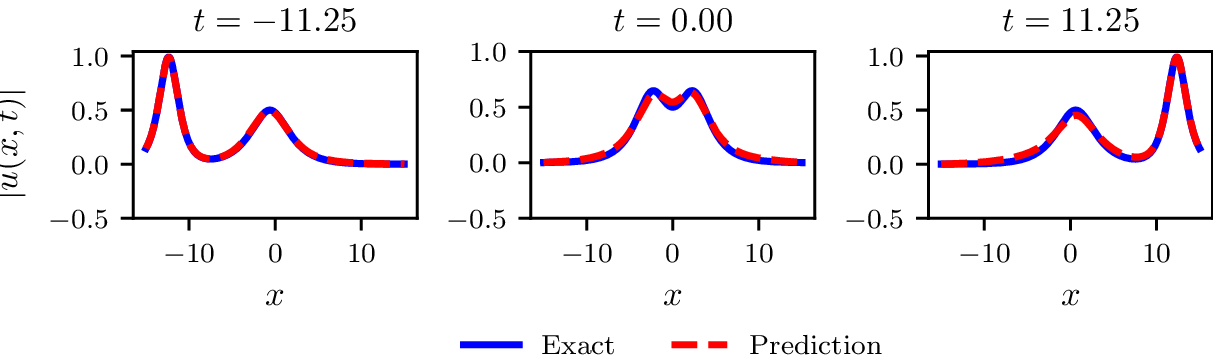}\\
\quad \quad \quad \quad  $(b)$\\
\includegraphics[width=7cm,height=5cm]{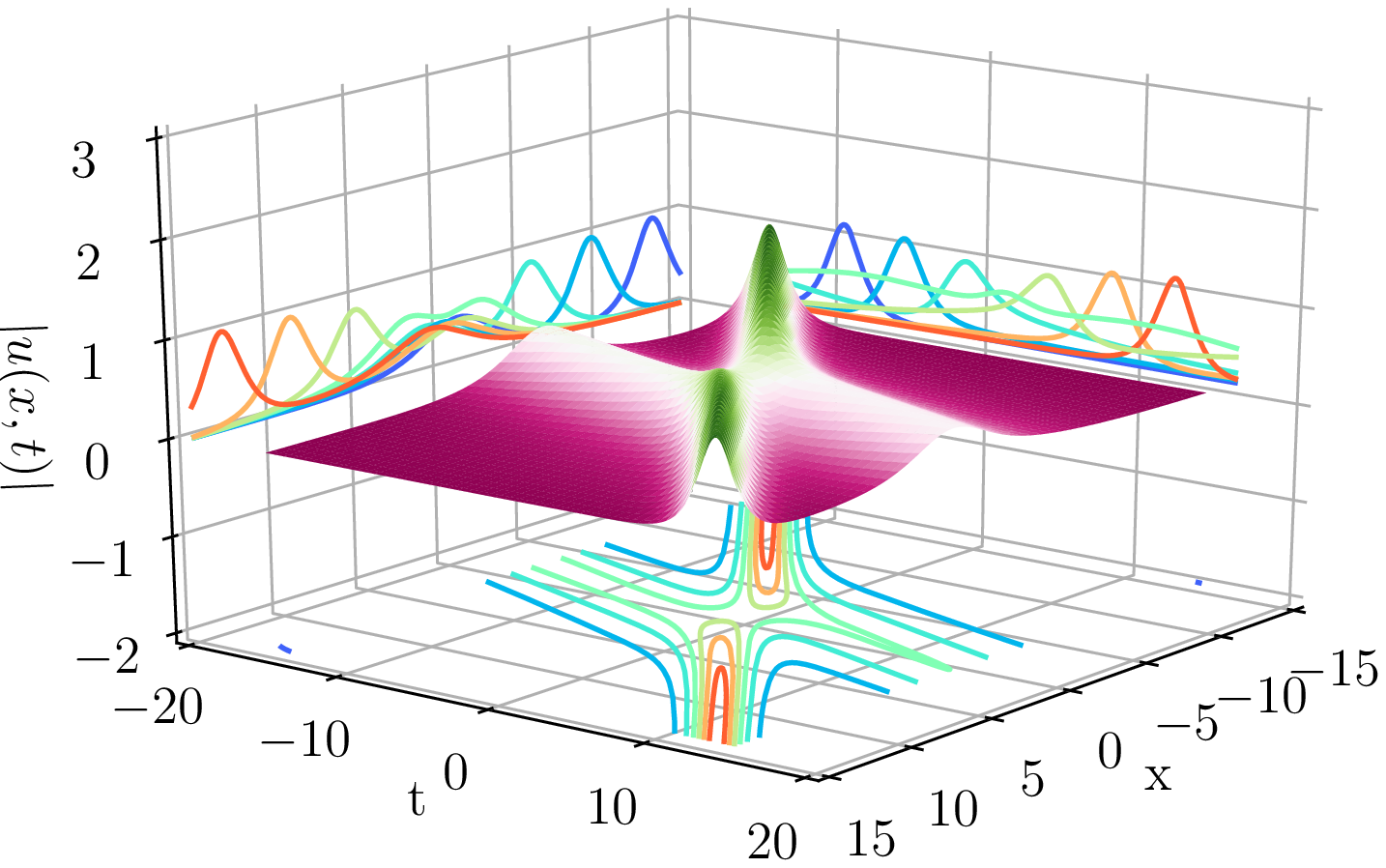}
\includegraphics[width=7cm,height=5cm]{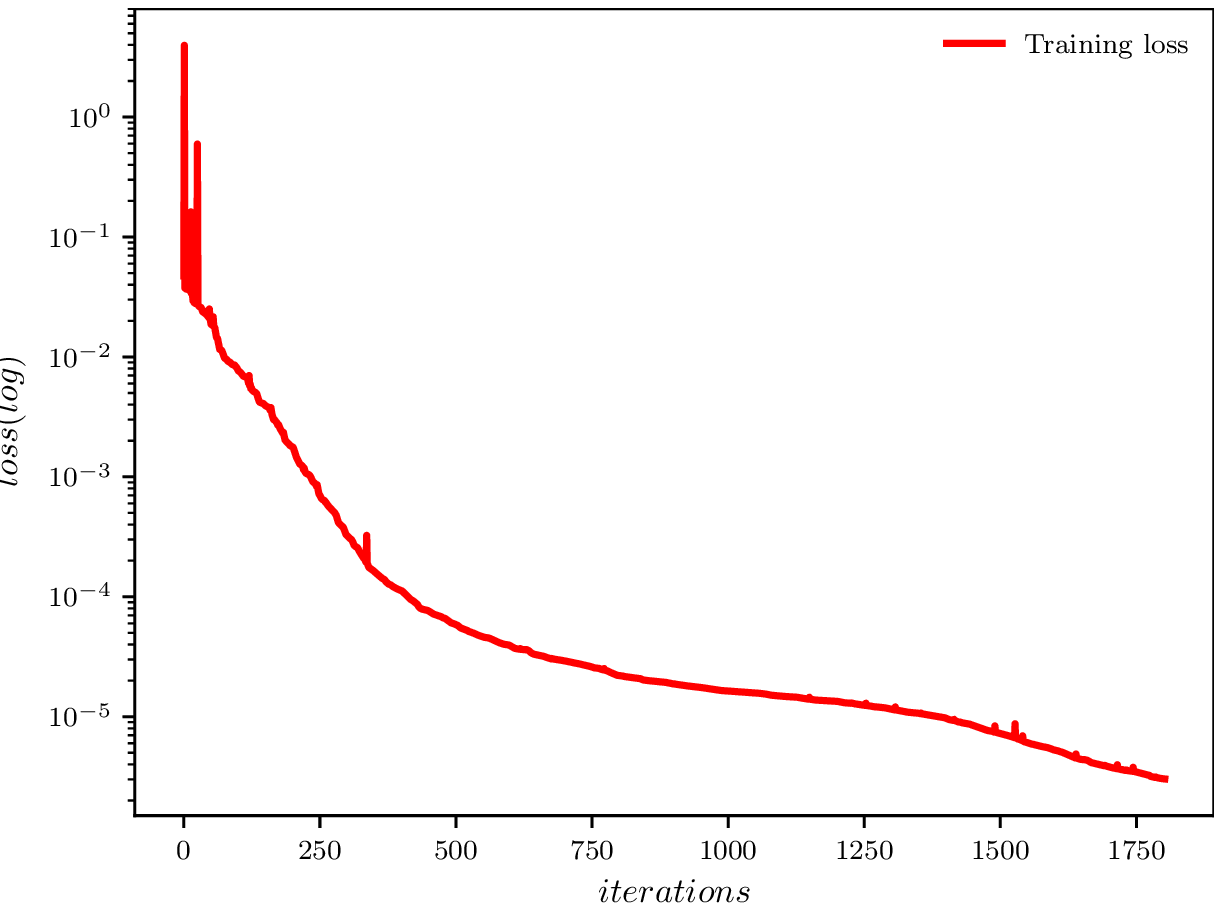}\\
$(c)$ \quad \quad \quad \quad \quad \quad \quad \quad \quad \quad \quad $(d)$
\caption{(Color online) The 2-soliton solution $u(x, t)$ for the nonlocal mKdV equation:(a) The density  plot and the  error density diagram;
(b) The wave propagation plot at three different times;
(c) The three-dimensional plot;
(d) The loss curve figure.}\label{3t}
\end{figure}

We select the same configuration points as the 1-soliton solution and use the LHS method to obtain training data sets, then we input these training data sets into the depth network of 9 hidden layers, each of which has 40 neurons. We have successfully learned the 2-soliton solution. Compared with the exact solution, its $\mathbb{L}_2$-norm  error is 5.937731e-02. The whole training time is 162.7119s, and the number of iterations is 1715. What needs to be noted here is not that more configuration points or more training network layers will lead to better results. Through experiments, we choose deeper network layers, and the results will be worse. The $\mathbb{L}_2$-norm  error will become 2.027299e-01. The training time is 155.5235s, and the number of iterations is 1402. The training results are shown in Fig.\ref{3t}, including the density diagram, error dynamics diagram, propagation diagram at different times, three-dimensional diagram and loss curve diagram of learning 2-soliton solution and accurate 2-soliton solution. It can be seen that the results are quite satisfactory.

\begin{figure}[htbp]
\centering
\includegraphics[width=12cm,height=3.5cm]{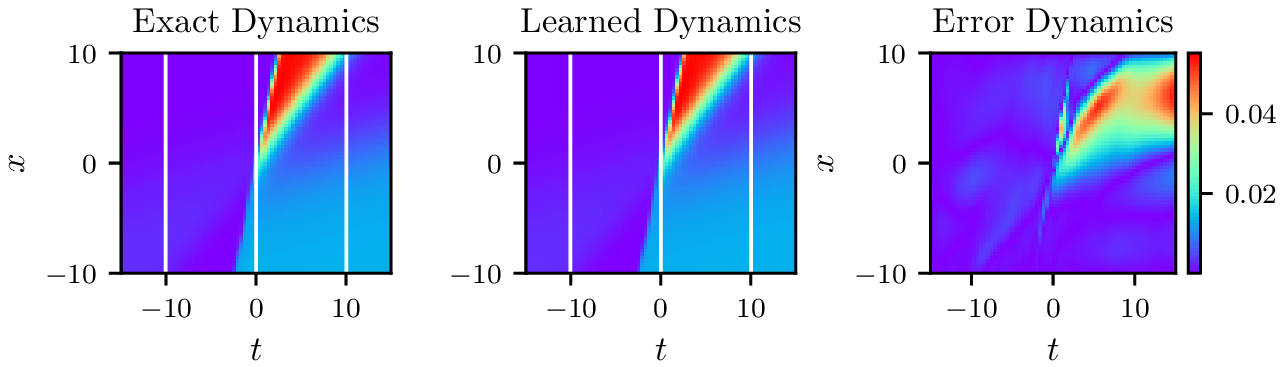}\\
\quad \quad \quad \quad   $(a)$\\
\includegraphics[width=12cm,height=4.5cm]{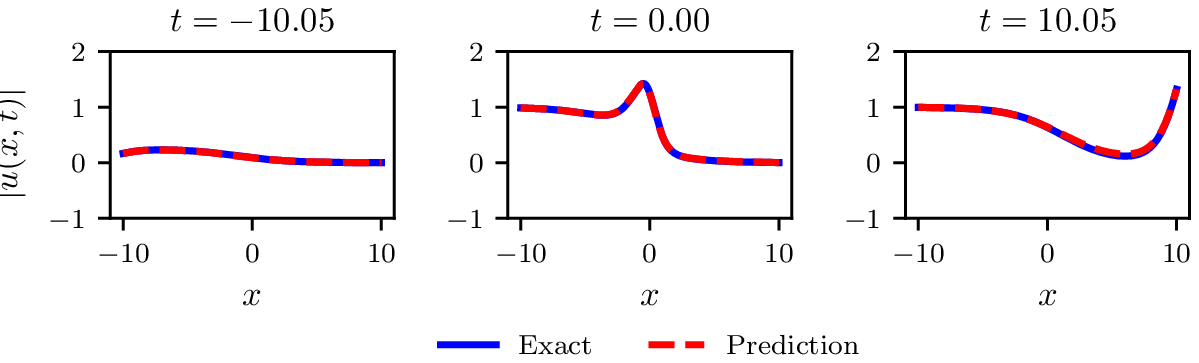}\\
\quad \quad \quad \quad  $(b)$\\
\includegraphics[width=7cm,height=5cm]{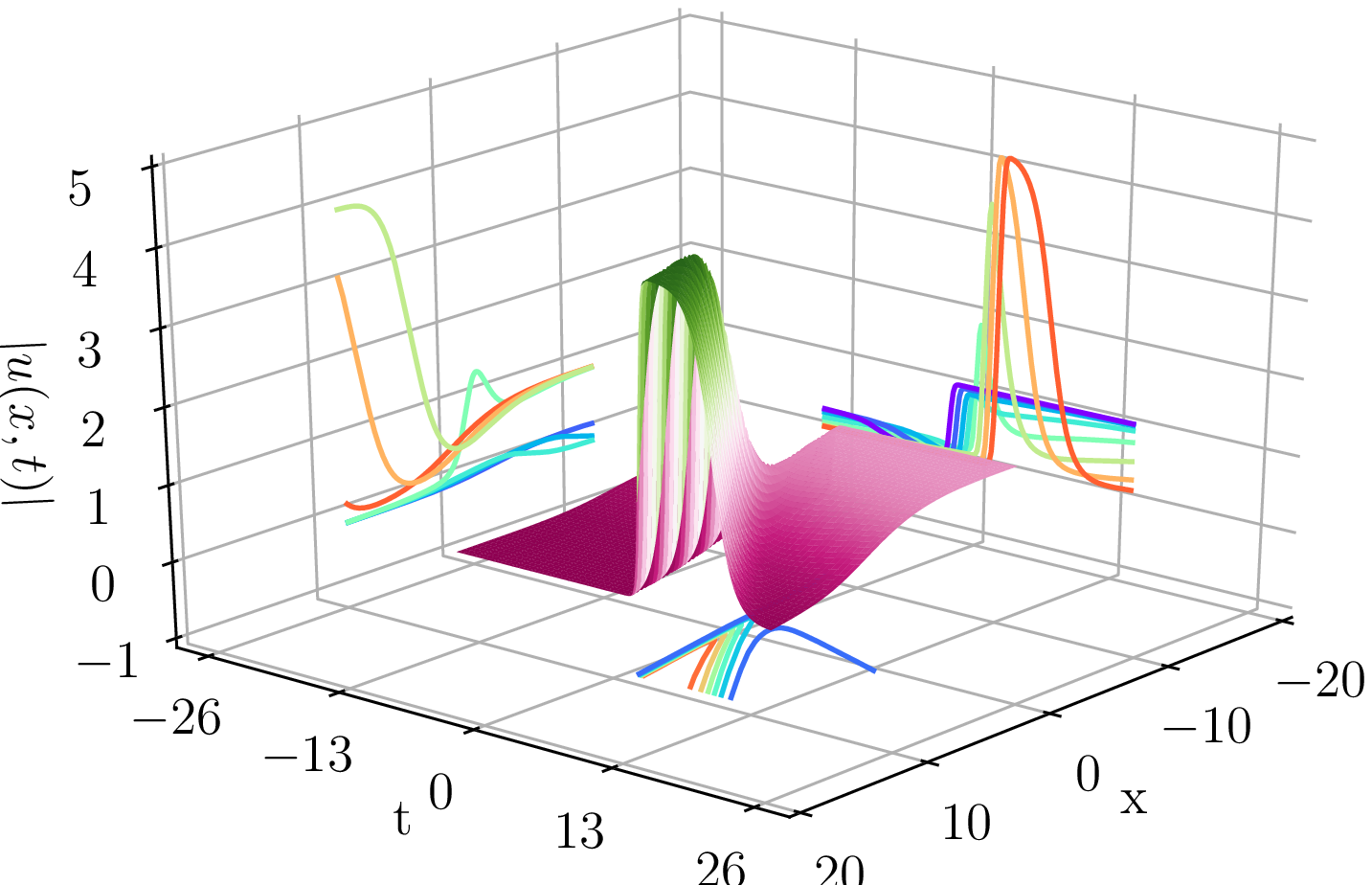}
\includegraphics[width=7cm,height=5cm]{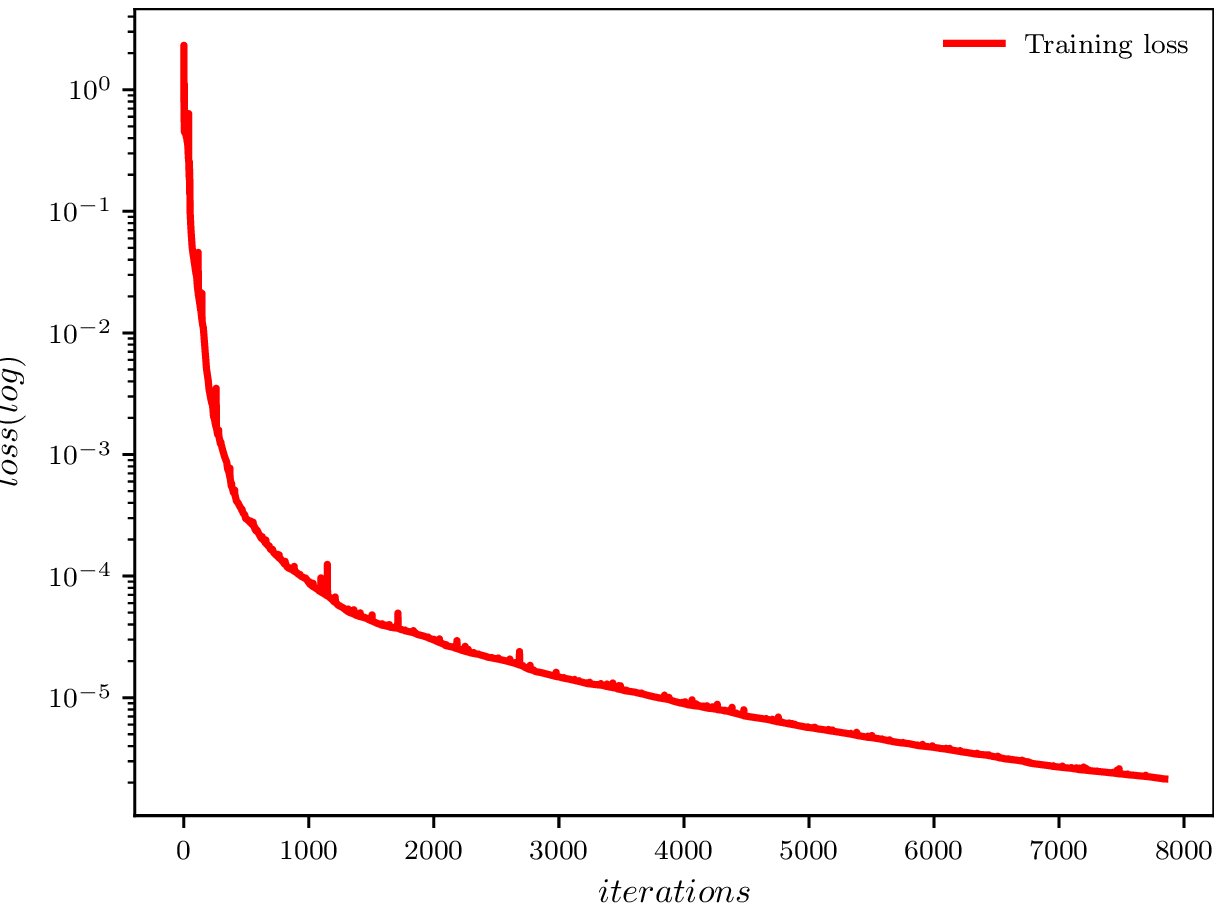}\\
$(c)$ \quad \quad \quad \quad \quad \quad \quad \quad \quad \quad \quad $(d)$
\caption{(Color online)  Solution of the interaction between soliton and kink-type waves $u(x, t)$ for the nonlocal mKdV equation:(a) The density  plot and the  error density diagram;
(b) The wave propagation plot at three different times;
(c) The three-dimensional plot;
(d) The loss curve figure.}\label{4t}
\end{figure}

When $k=0,l=2,a_1=a_2=b_1=b_2=-1$, Eq.(\ref{u20}) can be reduced to
\begin{equation}\label{2u}
u(x,t)==-\frac{1+\frac{9}{2} e^{x-t}+\frac{1}{2} e^{3 x-9 t}+4 e^{\frac{3}{2} x-\frac{9}{8} t}}{1+2 e^{3 x-9 t}+e^{\frac{3}{2} x-\frac{9}{8} t}+e^{-8 t+2 x}+2 e^{\frac{1}{2} x-\frac{1}{8} t}+e^{\frac{7}{2} x-\frac{73}{8} t}},
\end{equation}
which represents interaction of soliton and kink-type wave. The Dirichlet boundary $x$ and $t$ of Eq. (\ref{2u}) are [-10,10] and [-15,15] respectively. The initial training condition is
\begin{equation}
u(x,t)=-\frac{1+\frac{9}{2} e^{x+15}+\frac{1}{2} e^{3 x+135}+4 e^{\frac{3}{2} x+\frac{135}{8}}}{1+2 e^{3 x+135}+e^{\frac{3}{2} x+\frac{135}{8}}+e^{2 x+120}+2 e^{\frac{1}{2} x+\frac{15}{8}}+e^{\frac{7}{2} x+\frac{1095}{8}}}.
\end{equation}
Select the same configuration points as above to obtain training data, and put them into a 6-layer neural network with each 40 neurons. The solution of soliton kink interaction is well learned. Under the condition of training duration of 373.2689 seconds and iteration number of 7028, We get that the error of $\mathbb{L}_2$-norm between the exact solution and the training solution is 1.452126e-02. In Fig. \ref{4t}(a),(b), the density graph, error graph and time evolution graph of the training and exact solution at t=-10.05, t=0, t=10.05 are specifically given. In  Fig. \ref{4t}(c), (d), we use the training data to give the three-dimensional graph and the iterative degree graph.

\section{Data-driven solutions of nonlocal mKdV equation with nonzero boundary}
This part mainly studies the solutions of nonlocal mKdV equations under nonzero boundary, including kink solution, dark solution, anti-dark solution and rational solution.
\subsection{Data-driven solitary wave solution}
\begin{figure}[htbp]
\centering
\includegraphics[width=12cm,height=3.5cm]{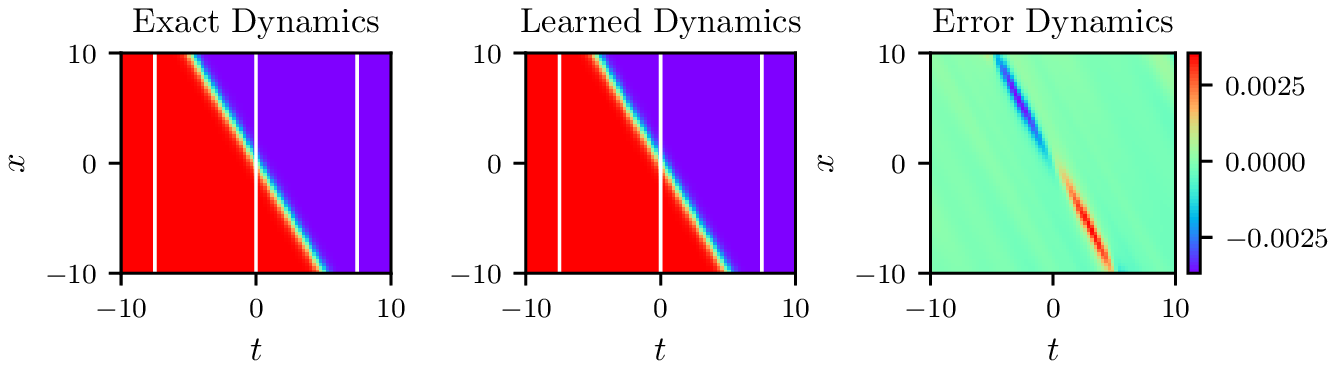}\\
\quad \quad \quad \quad   $(a)$\\
\includegraphics[width=12cm,height=4.5cm]{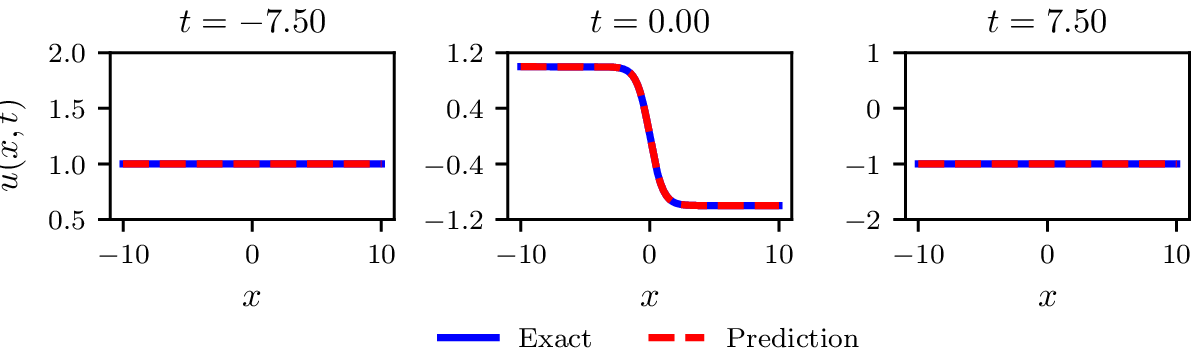}\\
\quad \quad \quad \quad  $(b)$\\
\includegraphics[width=7cm,height=5cm]{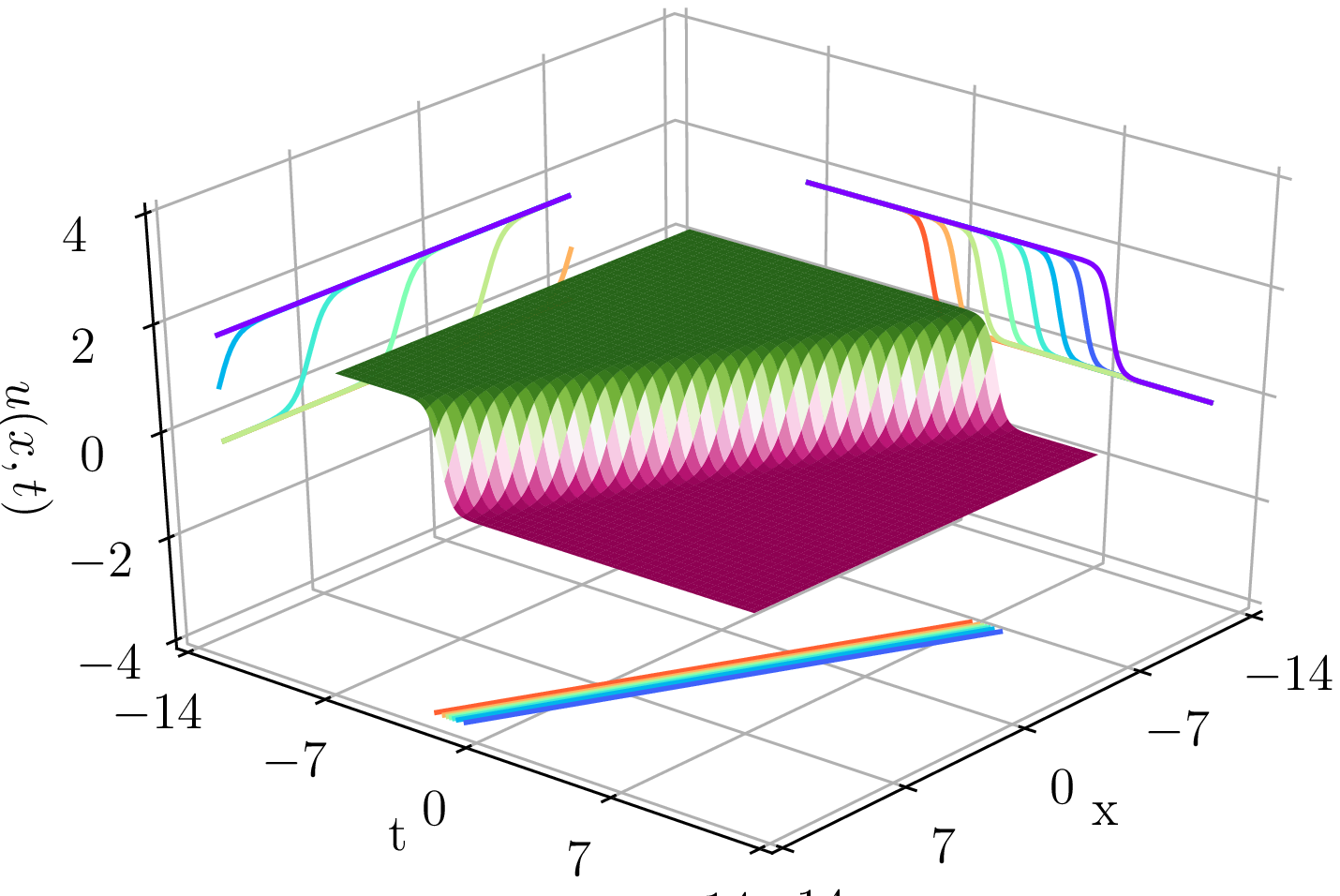}
\includegraphics[width=7cm,height=5cm]{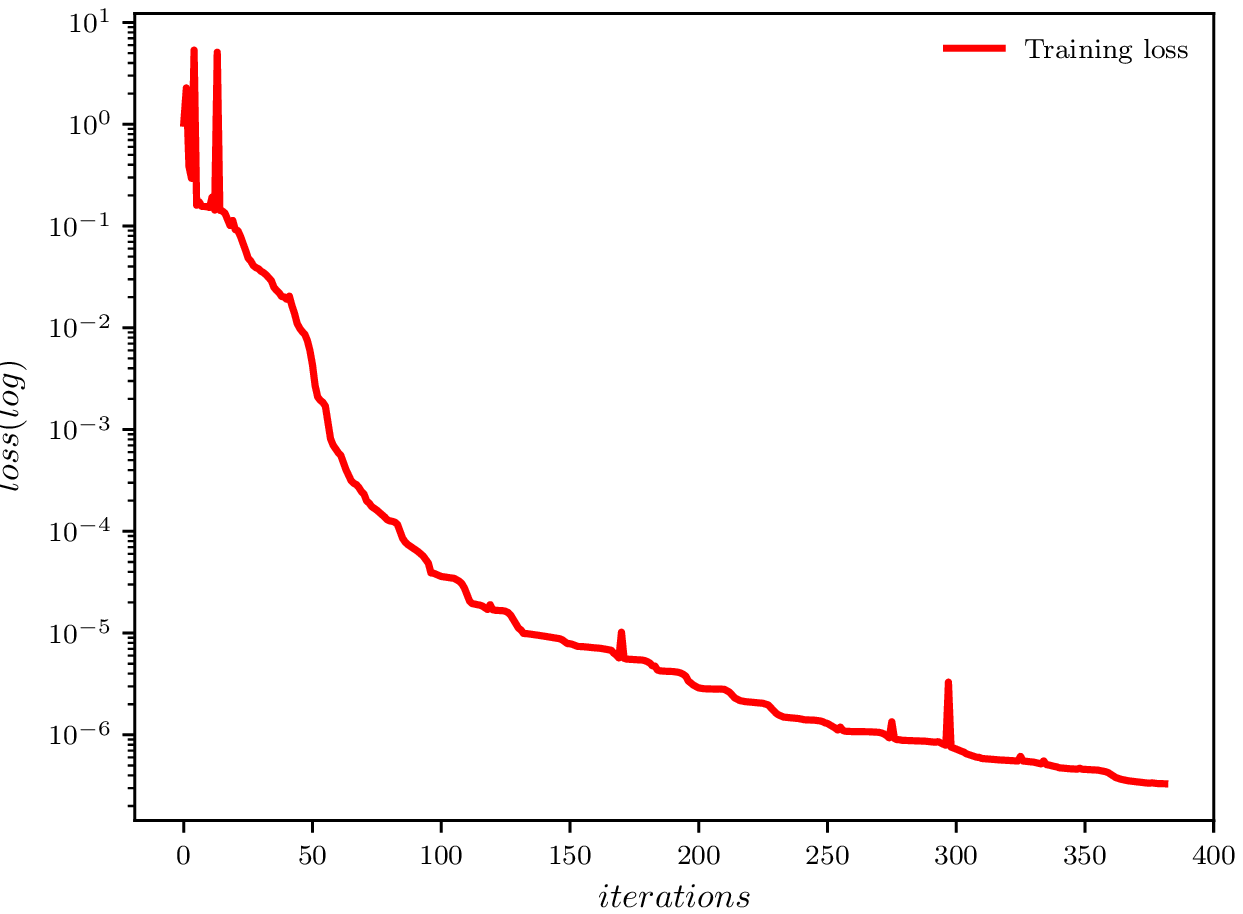}\\
$(c)$ \quad \quad \quad \quad \quad \quad \quad \quad \quad \quad \quad $(d)$
\caption{(Color online) The kink soliton solution $u(x, t)$ for the nonlocal mKdV equation under nonzero boundary:(a) The density  plot and the  error density diagram;
(b) The wave propagation plot at three different times;
(c) The three-dimensional plot;
(d) The loss curve figure.}\label{5t}
\end{figure}

The kink solution of nonlocal mKdV with non-zero boundary has been obtained by inverse scattering method in Ref.\cite{ZG-2020-PD}, which is expressed as
$$
u(x, t)=-\tilde{u}_0\tanh \left[\tilde{u}_0\left(x+2 \tilde{u}_0^2 t\right)+\frac{\theta_1+\theta_{2}}{2} i\right] e^{i \theta_{2}}.
$$
In order to distinguish between the training initial boundary and the non-zero boundary conditions satisfied by the equation, we use $\tilde{u}_0$ to represent the non-zero boundary conditions satisfied by the equation.
 The neural network is constructed by Eq. (\ref{f}), and its training initial condition is
$$
u_0(x)=u(x,-10)=-\tilde{u}_0\tanh \left[\tilde{u}_0\left(x-20 \tilde{u}_0^2\right)+\frac{\theta_1+\theta_{2}}{2} i\right] e^{i \theta_{2}}.
$$
At this time, we select the region of $x$ and $t$ as $[-10,10]$, and divide 512 points and 200 points in the region of $x$ and $t$ respectively. $N_u=1000$ and $N_f=10000$ points are randomly selected as training data at the boundary and inside respectively. We numerically predict the kink soliton solution with proper parameters $\tilde{u}_0=1, \theta_1=\theta_2=0$ under a 6-hidden-layer deep PINN with the 40 neurons per layer. The training results show that its $\mathbb{L}_2$ error is 5.162946e-04 compared with the exact solution, the training time is merely 23.7374s, and the number of iterations is only 350, the final result is shown in Fig. \ref{5t}. Once again, it shows the power of the deep integrable system.

\begin{figure}[htbp]
\centering
\includegraphics[width=12cm,height=3.5cm]{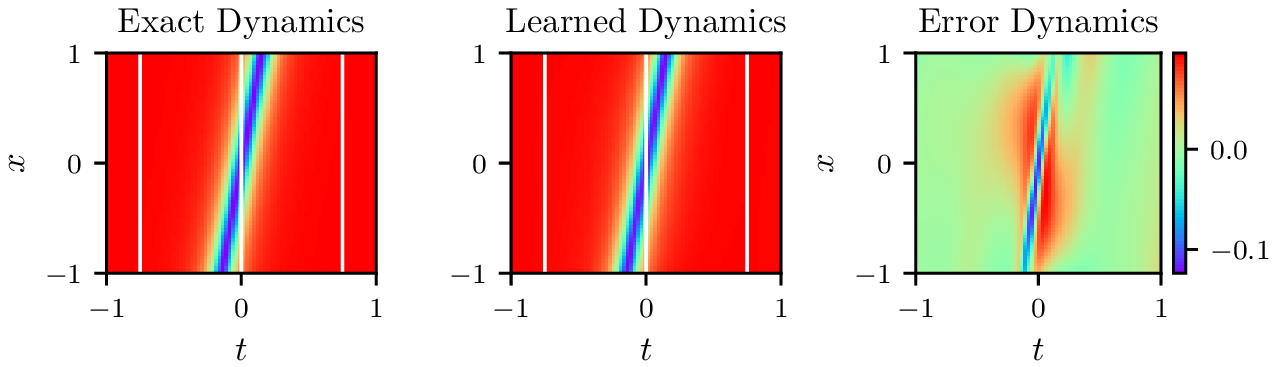}\\
\quad \quad \quad \quad   $(a)$\\
\includegraphics[width=12cm,height=4.5cm]{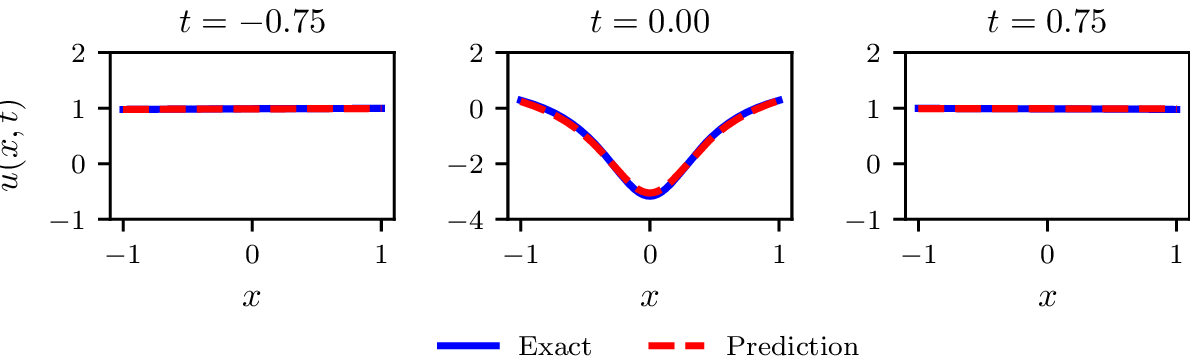}\\
\quad \quad \quad \quad  $(b)$\\
\includegraphics[width=7cm,height=5cm]{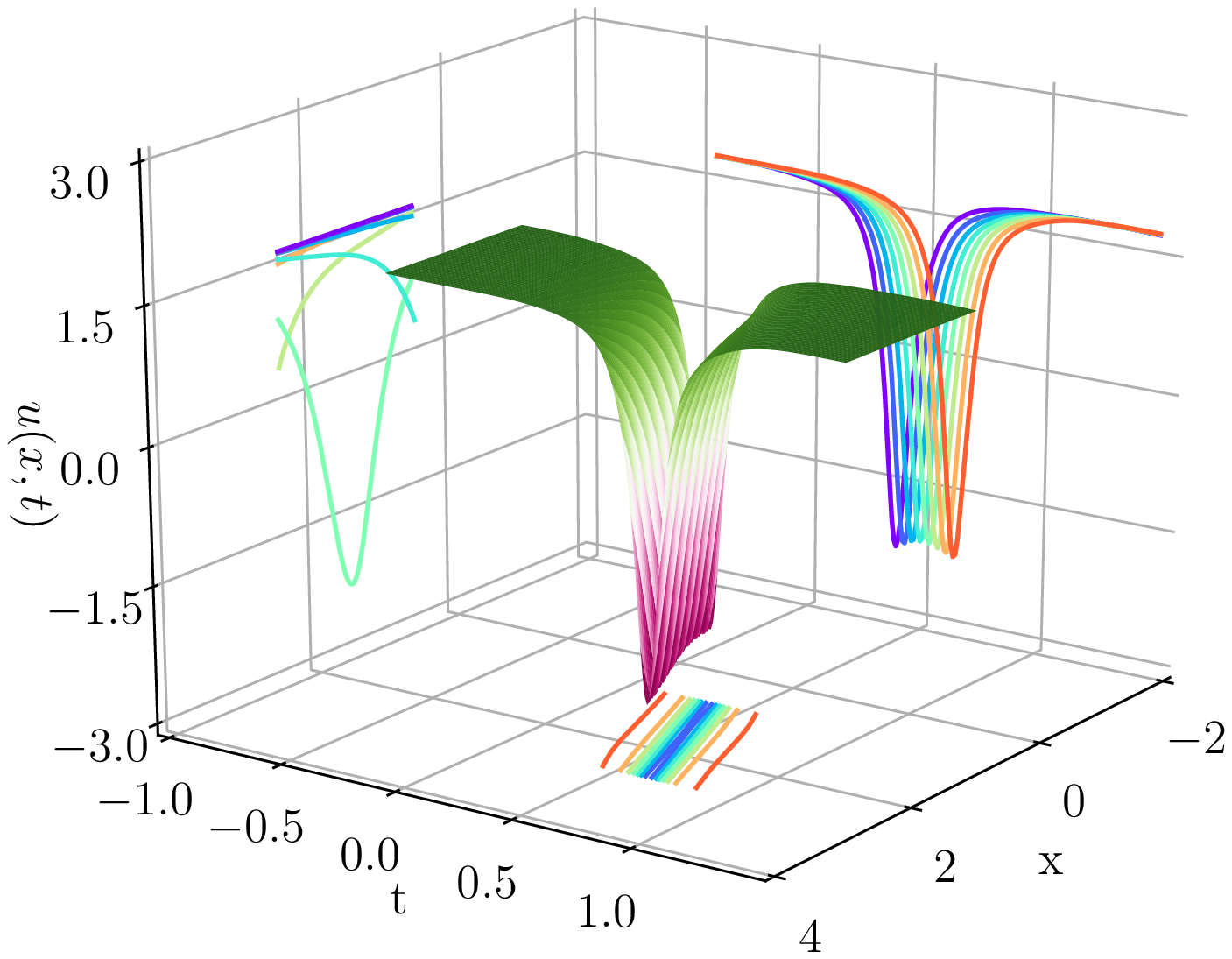}
\includegraphics[width=7cm,height=5cm]{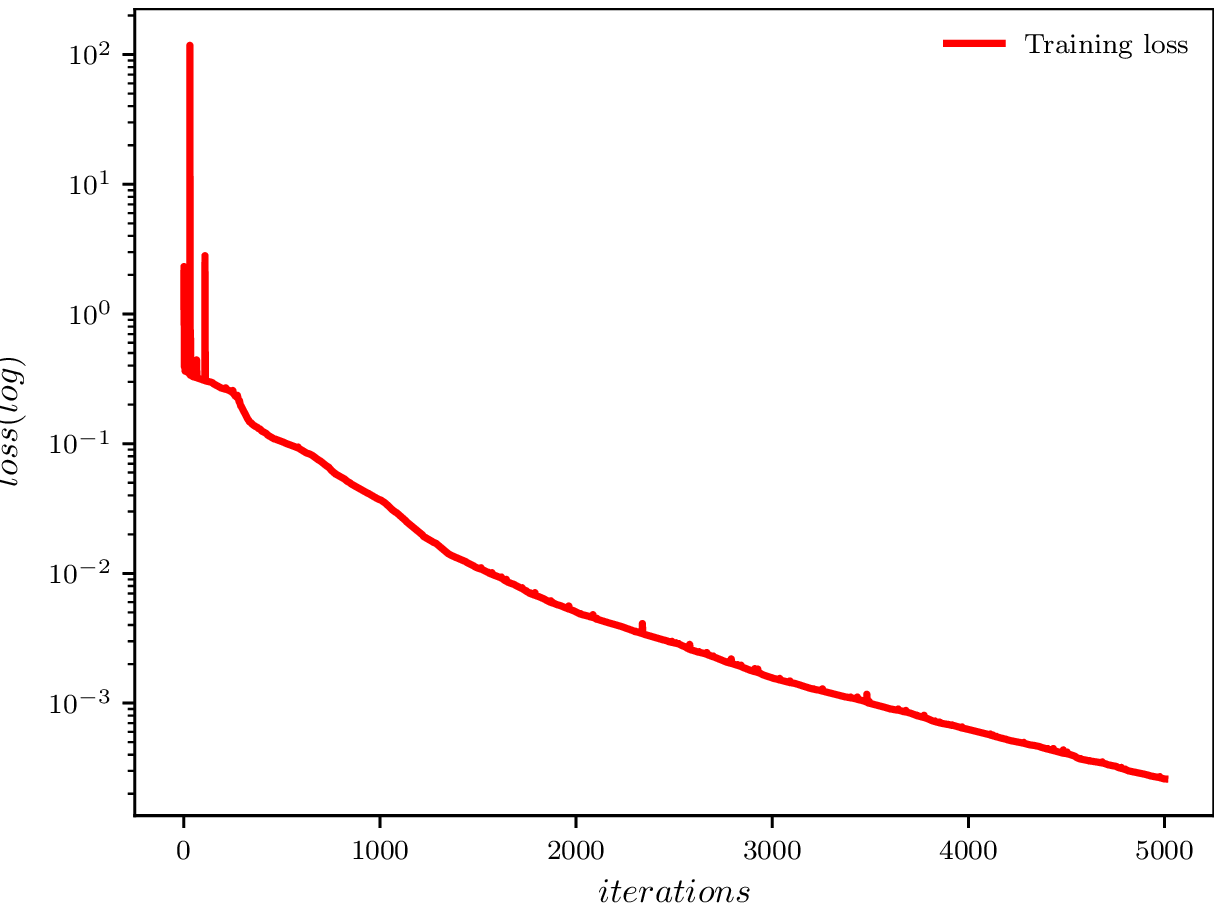}\\
$(c)$ \quad \quad \quad \quad \quad \quad \quad \quad \quad \quad \quad $(d)$
\caption{(Color online) The dark soliton solution $u(x, t)$ for the nonlocal mKdV equation:(a) The density  plot and the  error density diagram;
(b) The wave propagation plot at three different times;
(c) The three-dimensional plot;
(d) The loss curve figure.}\label{6t}
\end{figure}

\begin{figure}[htbp]
\centering
\includegraphics[width=12cm,height=3.5cm]{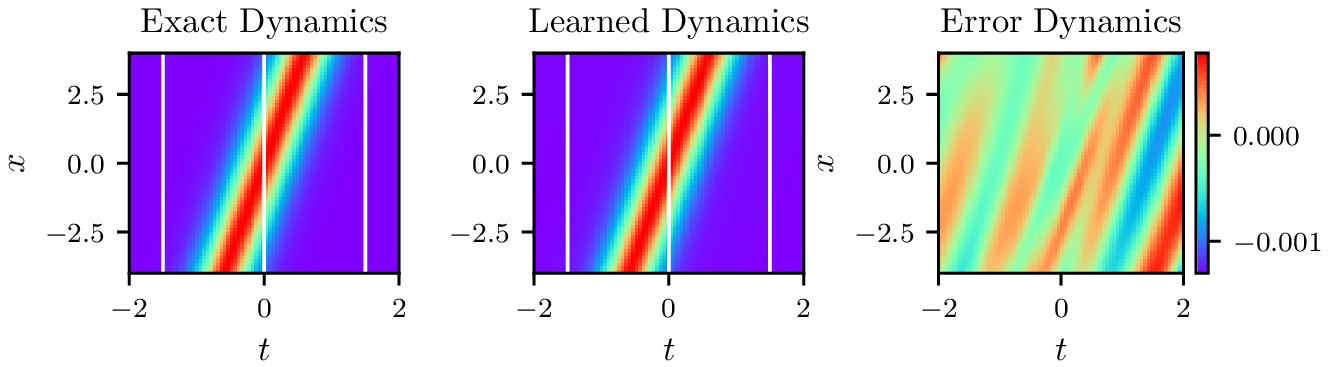}\\
\quad \quad \quad \quad   $(a)$\\
\includegraphics[width=12cm,height=4.5cm]{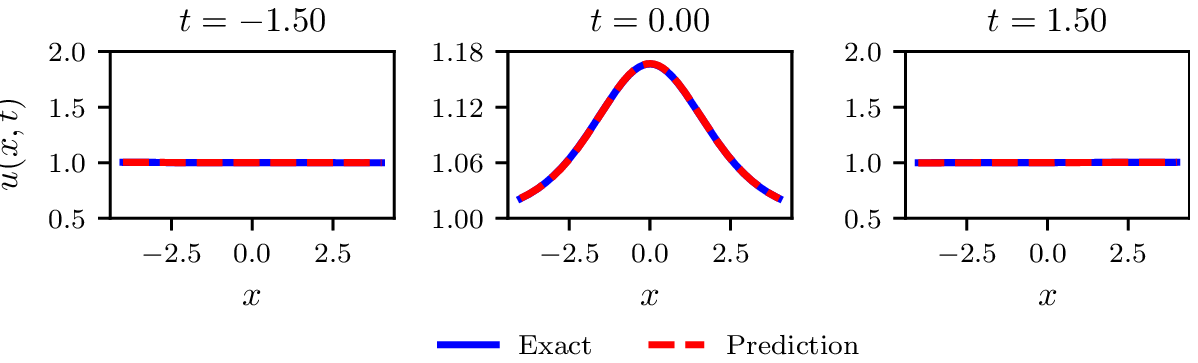}\\
\quad \quad \quad \quad  $(b)$\\
\includegraphics[width=7cm,height=5cm]{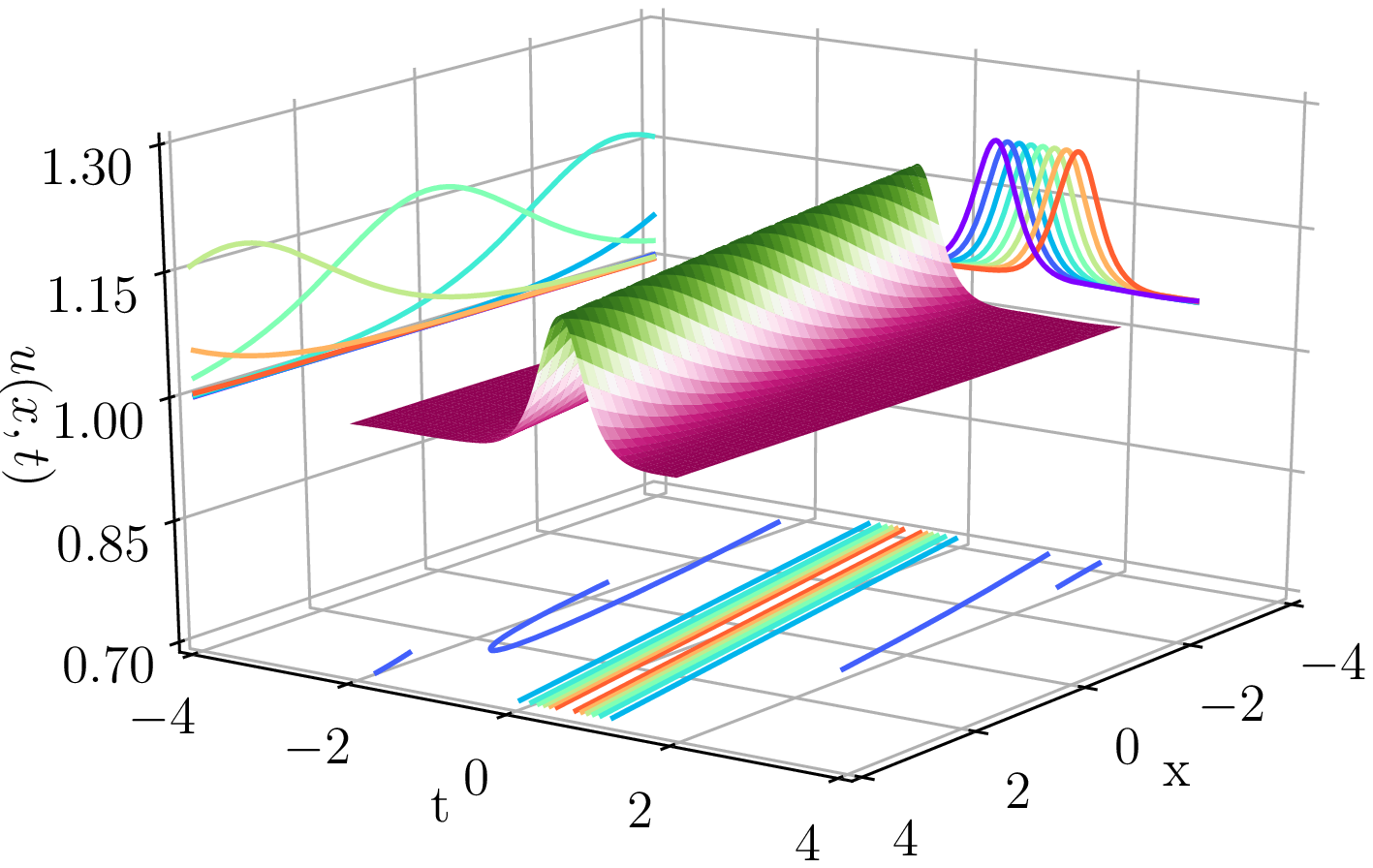}
\includegraphics[width=7cm,height=5cm]{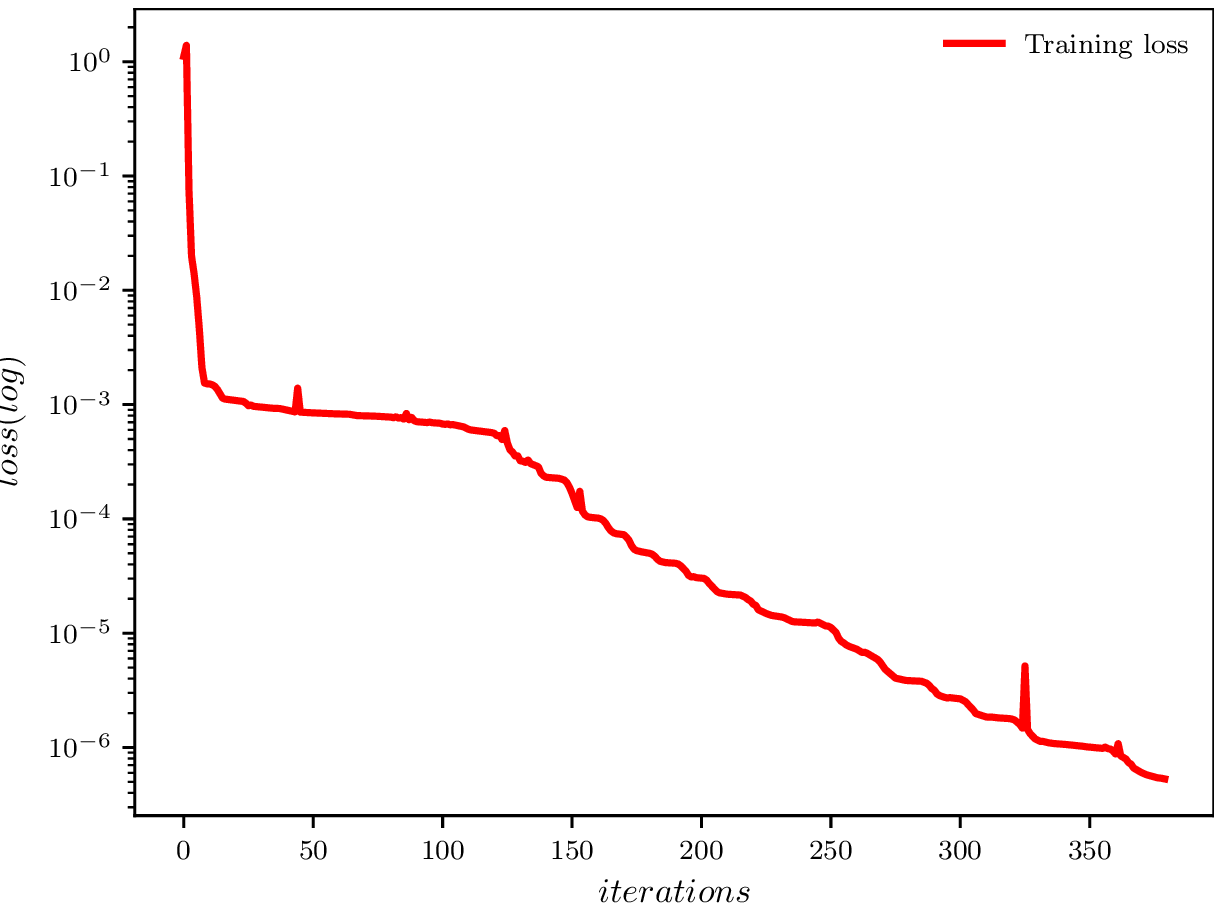}\\
$(c)$ \quad \quad \quad \quad \quad \quad \quad \quad \quad \quad \quad $(d)$
\caption{(Color online) The anti-dark soliton solution $u(x, t)$ for nonlocal mKdV equation:(a) The density  plot and the  error density diagram;
(b) The wave propagation plot at three different times;
(c) The three-dimensional plot;
(d) The loss curve figure.}\label{7t}
\end{figure}

The dark and anti-dark soliton solutions of the nonlocal mKdV are also given in Ref. \cite{ZG-2020-PD}, and the overall expression is
\begin{equation}
\begin{gathered}
u(x, t)=\frac{e^{i \theta_{-}}}{w} \frac{w \tilde{u}_0\left(\tilde{u}_0^2+w^2\right)\left[e^{2 \varphi+i\left(\theta_1+\theta_2\right)}-e^{2 i \theta_{-}}\right]+2\left(w^4 e^{i \theta_2}-\tilde{u}_0^4 e^{i \theta_1}\right) e^{\varphi+i \theta_{-}}}{\left(\tilde{u}_0^2+w^2\right)\left[e^{2 \varphi+i\left(\theta_1+\theta_2\right)}-e^{2 i \theta_{-}}\right]+2 \tilde{u}_0 w\left(e^{i \theta_2}-e^{i \theta_1}\right) e^{\varphi+i \theta_{-}}},
\end{gathered}
\end{equation}
where
$$
\varphi=\frac{\left(w^2-\tilde{u}_0^2\right)\left[w^2 x-\left(w^4+4 w^2 \tilde{u}_0^2+\tilde{u}_0^4\right) t\right]}{w^3} .
$$
it is a dark soliton solution as $\tilde{u}_0=1,w=\frac{3}{2}, \theta_1=\pi, \theta_2=\theta=0,$ and it is an anti-dark soliton solution as $\tilde{u}_0=1,w=\frac{3}{2}, \theta_1=\theta_=0,\theta_2=\pi$. For the dark soliton solution, let the interval of $x$ and $t$ be [-1,1], and the corresponding initial condition is
$$
u_0(x)=u(x,-1)=\frac{39 e^{\frac{5}{3} x+\frac{1205}{108}}+39-97 e^{\frac{5}{6} x+\frac{1205}{216}}}{39 e^{\frac{5}{3} x+\frac{1205}{108}}+39-72 e^{\frac{5}{6} x+\frac{1205}{216}}}.
$$
For the anti-dark soliton solution, let the intervals of $x$ and $t$ as [-4,4] and [-2,2] respectively, and its initial condition becomes
$$
u_0(x)=u(x,-2)=\frac{39 e^{\frac{5}{3} x+\frac{1205}{54}}+39+97 e^{\frac{5}{6} x+\frac{1205}{108}}}{39 e^{\frac{5}{3} x+\frac{1205}{54}}+39+72 e^{\frac{5}{6} x+\frac{1205}{108}}}.
$$
By performing the same data collection and training procedures as kink soliton solution, through training, we can get that the $\mathbb{L}_2$-norm error between the learning solution and the accurate solution is 2.459108e-02 for the dark soliton solution, the whole learning process takes about 261.6747 seconds, and iterates 4618. For the anti-dark soliton solution, the $\mathbb{L}_2$-norm error between the learning solution and the exact solution is 3.244125e-04, and the whole learning process takes about 24.5986 seconds, with 345 iterations. The construction of this neural network is easier to learn the anti-dark soliton solution, takes less time, and the training result is more ideal.
The learning results are shown in Fig. \ref{6t} and Fig. \ref{7t} respectively.

\subsection{Data-driven rational solution}
\begin{figure}[htbp]
\centering
\includegraphics[width=12cm,height=3.5cm]{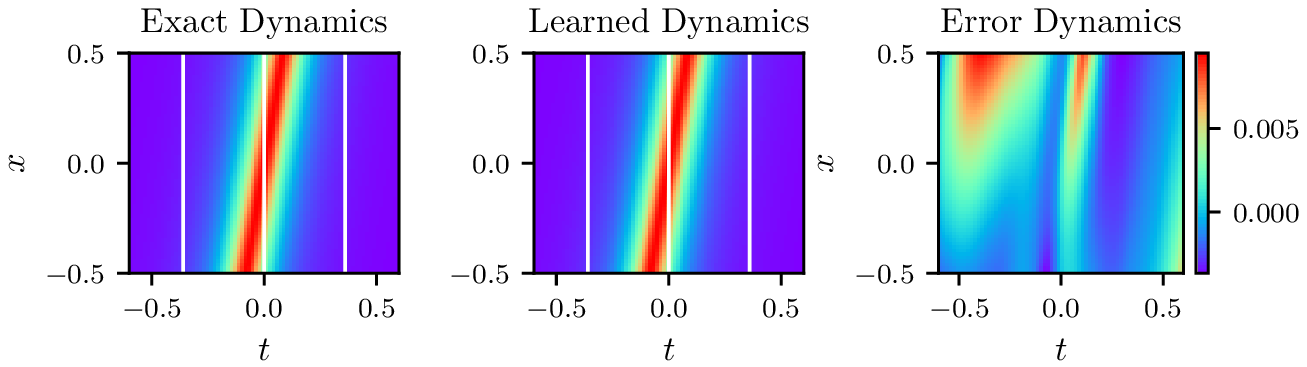}\\
\quad \quad \quad \quad   $(a)$\\
\includegraphics[width=12cm,height=4.5cm]{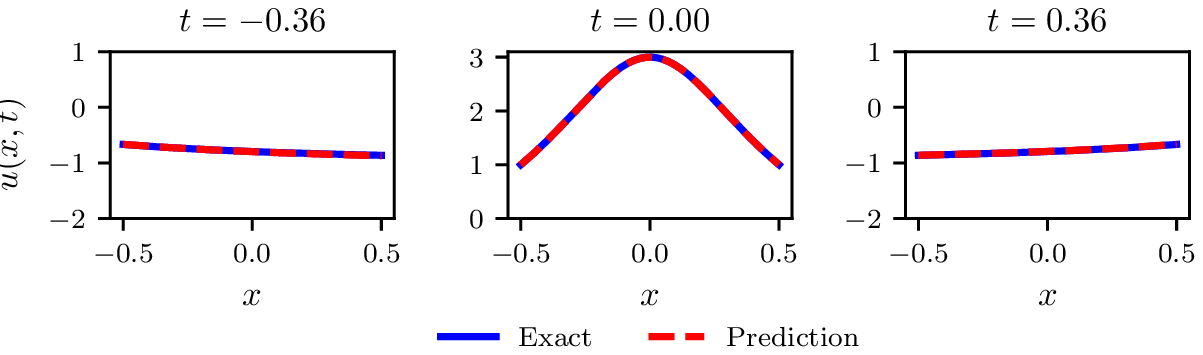}\\
\quad \quad \quad \quad  $(b)$\\
\includegraphics[width=7cm,height=5cm]{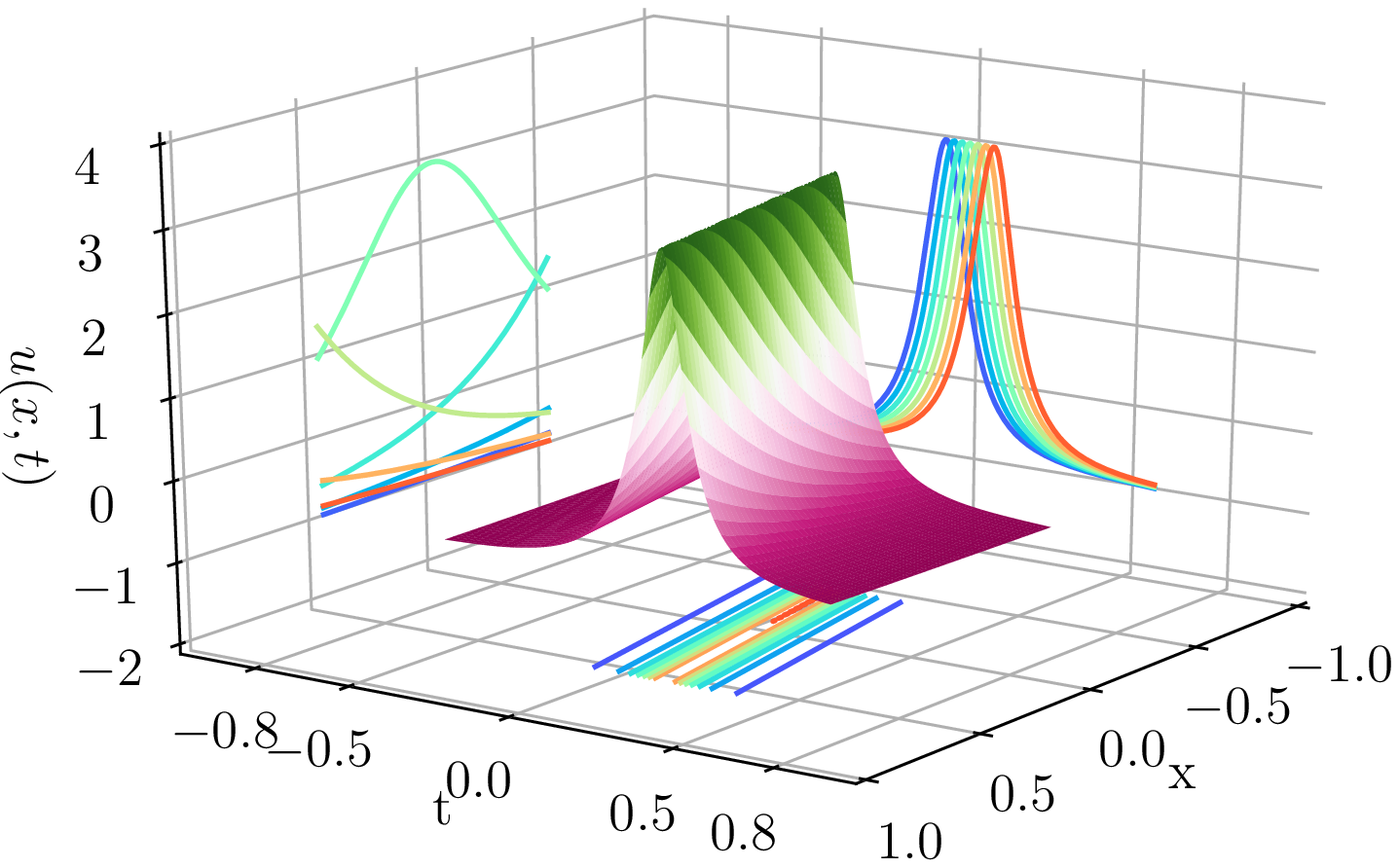}
\includegraphics[width=7cm,height=5cm]{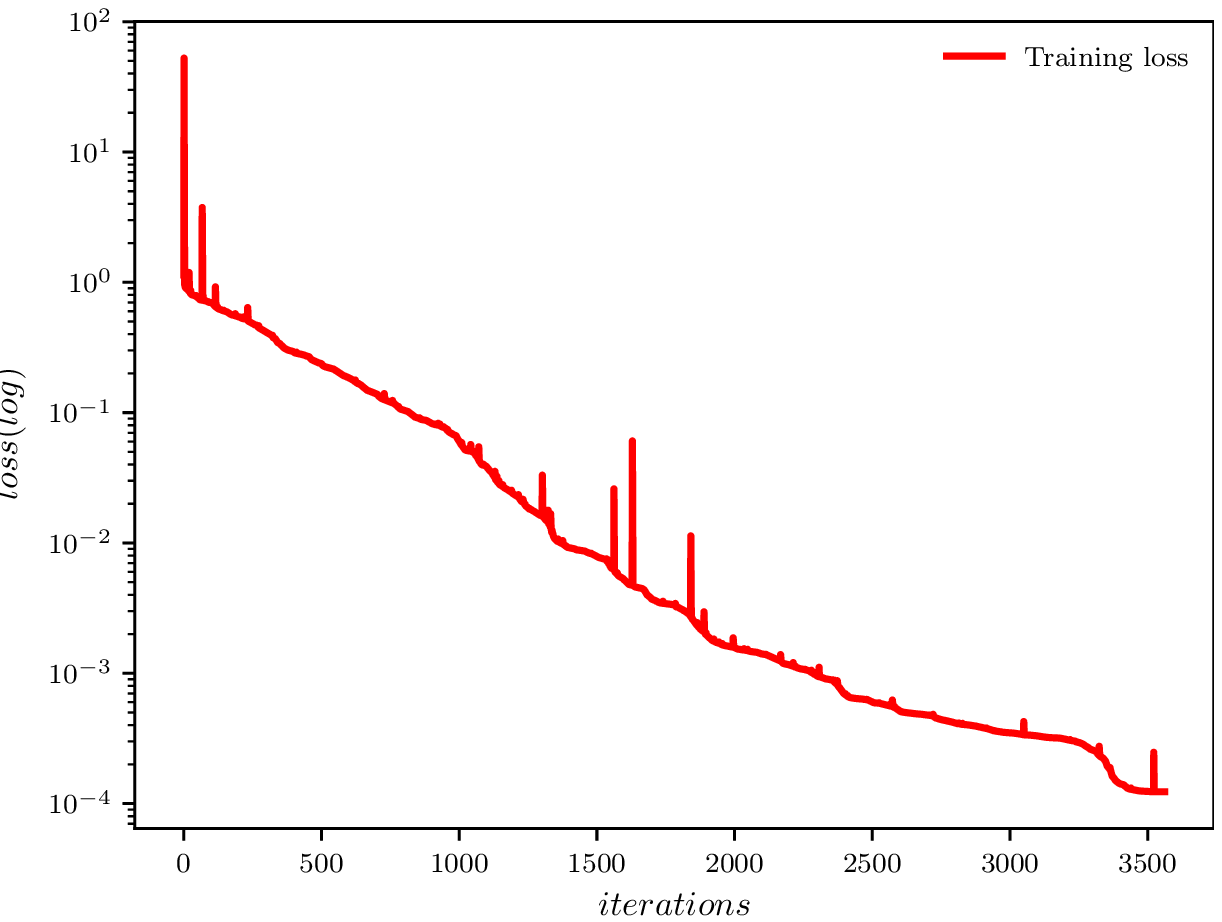}\\
$(c)$ \quad \quad \quad \quad \quad \quad \quad \quad \quad \quad \quad $(d)$
\caption{(Color online)  The rational soliton solution $u(x, t)$ for the nonlocal mKdV equation:(a) The density  plot and the  error density diagram;
(b) The wave propagation plot at three different times;
(c) The three-dimensional plot;
(d) The loss curve figure.}\label{8t}
\end{figure}

In this section, we use the PINN method to construct the rational soliton solution of Eq. (15). The form of its solution is given in Ref.\cite{ZG-2020-PD}
\begin{equation}\label{uyl}
u(x,t)=-1+\frac{4}{1+4(x-6 t)^2}.
\end{equation}
If the boundary regions of $x$ and $t$ are [-0.5,0.5] and [-0.3,0.3], then the corresponding initial condition is
 \begin{equation}
u(x,t)=-1+\frac{4}{1+4(x-\frac{5}{9})^2},
\end{equation}
in order to learn rational soliton solution better and more accurately, $N_u=100$ configuration points are selected at the boundary and $N_f=5000$ configuration points are selected internally for training. The training results are compared with the accurate solution achieve a relative $\mathbb{L}_2$ error of 1.164655e-02 in about 477.8897s, and the number of iterations is 2848. In Fig. \ref{8t}, we give the specific training results, including the density graph, error graph and time evolution graph of the exact solution and the training solution. The 3D stereoscopic map is generated from the training results. The training error map is not very stable, and there will be some burrs in the middle, but this will not affect the final training results.

\section{ The PINN algorithm for the data-driven parameter discovery}
In this section, we focus on data driven discovery of the nonlocal mKdV equation through PINN algorithm. It is well known that linear equations have good properties, while most equations in real life are nonlinear. It is precisely because of the existence of these nonlinear terms that more physical properties are stimulated and widely used in various fields, such as physics, mechanics, life sciences, etc. Therefore, it is necessary to learn the specific forms of nonlinear terms. For nonlocal nonlinear partial differential mkdv equation, we mainly study the coefficient $a$ of the following nonlinear terms
\begin{equation}\label{amkdv}
u_{xxx}+u_{t}+auu(-x,-t)u_{x}=0,
\end{equation}
Theoretically, parameter can be found by using any known nonlocal mKdV solution. Here, we choose to use the complex solution to do parameter discovery, and choose Eq. (\ref{fpf}) and Eq.(\ref{fqf}) as the physical information neural network of the nonlocal mKdV equation. The Latin hypercube is still used for sampling. With the help of the exact soliton solutions of a=6 and $(x, t)\in [-5, 5]\times[-\frac{1}{100}, \frac{1}{100}]$, the training data set is generated by randomly selecting $N_u=200$ as the initial boundary data and $N_f=50000$ as the configuration point. According to the training data set obtained, the data-driven parameter $a$ can be found by using the PINN.

When the hidden layer of PINN is determined, we give the data-driven parameter a under different neurons of each layer, and show the results in Table \ref{tableA}. From the table, it can be found that, when no noise is added to the system, the precision of PINN learning unknown parameters becomes higher and higher as the number of neurons increases. When we add different noises, the results of parameter discovery will oscillate with the increase of the number of neurons. When the number of hidden layers and neurons is fixed, the number of internal configuration points also affects the discovery of parameters. In Table \ref{table2}, we list the results of parameter learning under different internal configuration points and different noises. It can be seen from the table that the more internal configuration points are given, the better the learning results will be. However, when the internal configuration points are determined, different noises are added to the neural network, and the results are also oscillatory. In general, the more configuration points, the greater the noise added to the network, and the result of parameter discovery is the closest to the ideal result.

\begin{figure}[htbp]
\centering
\includegraphics[width=7cm,height=5cm]{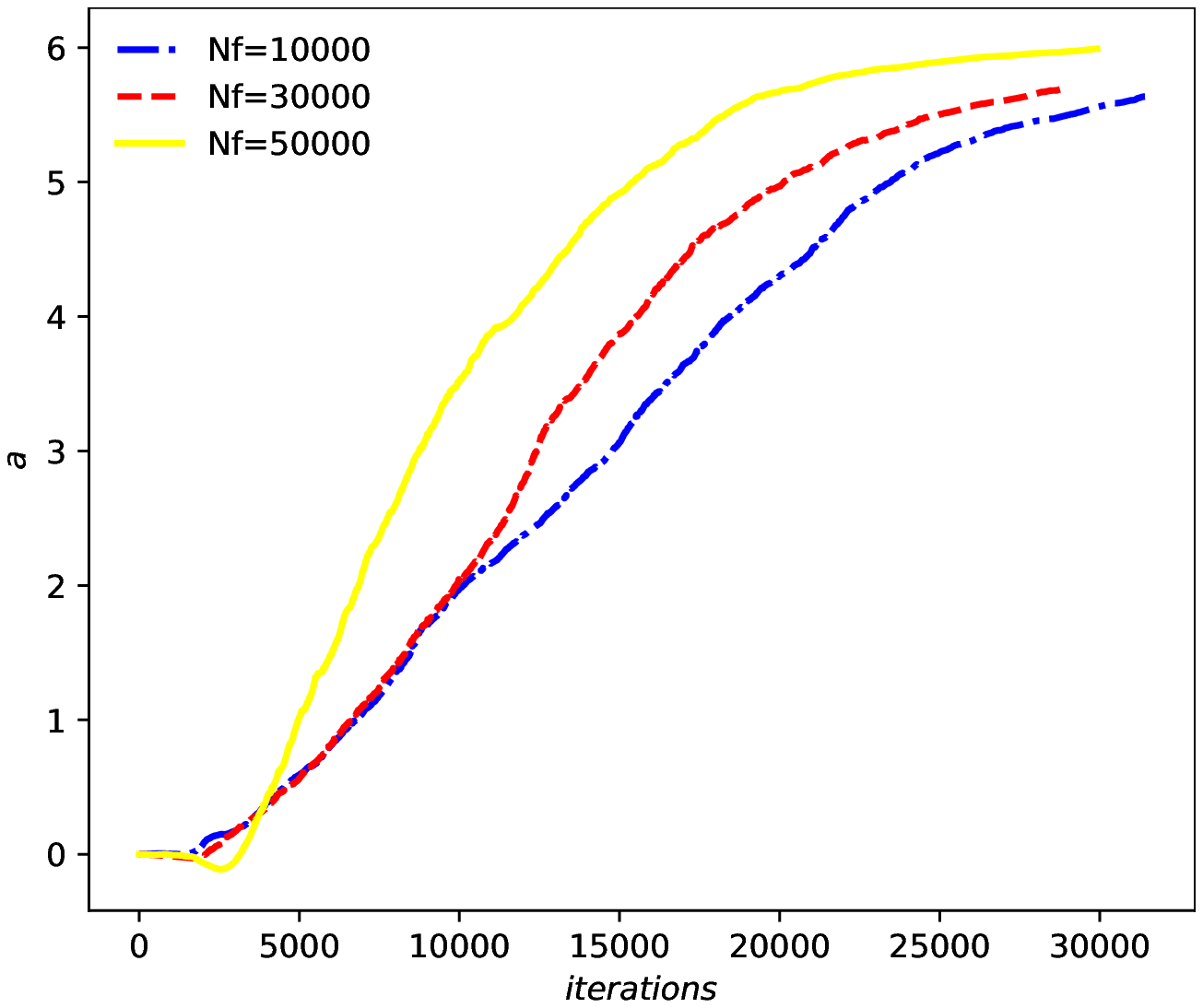}
\includegraphics[width=7cm,height=5cm]{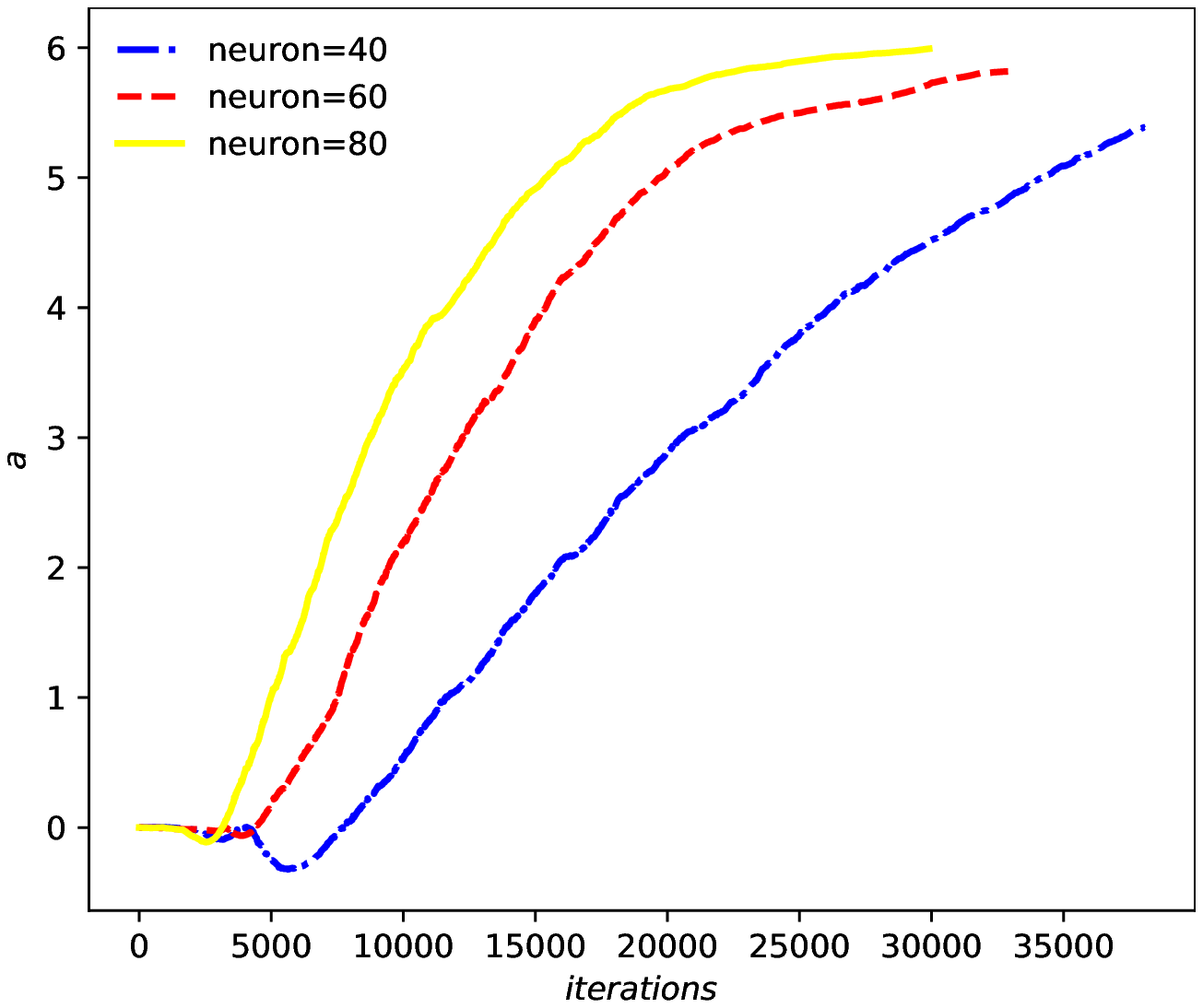}\\
$(a)$ \quad \quad \quad \quad \quad \quad \quad \quad \quad \quad \quad $(b)$\\
\includegraphics[width=7cm,height=5cm]{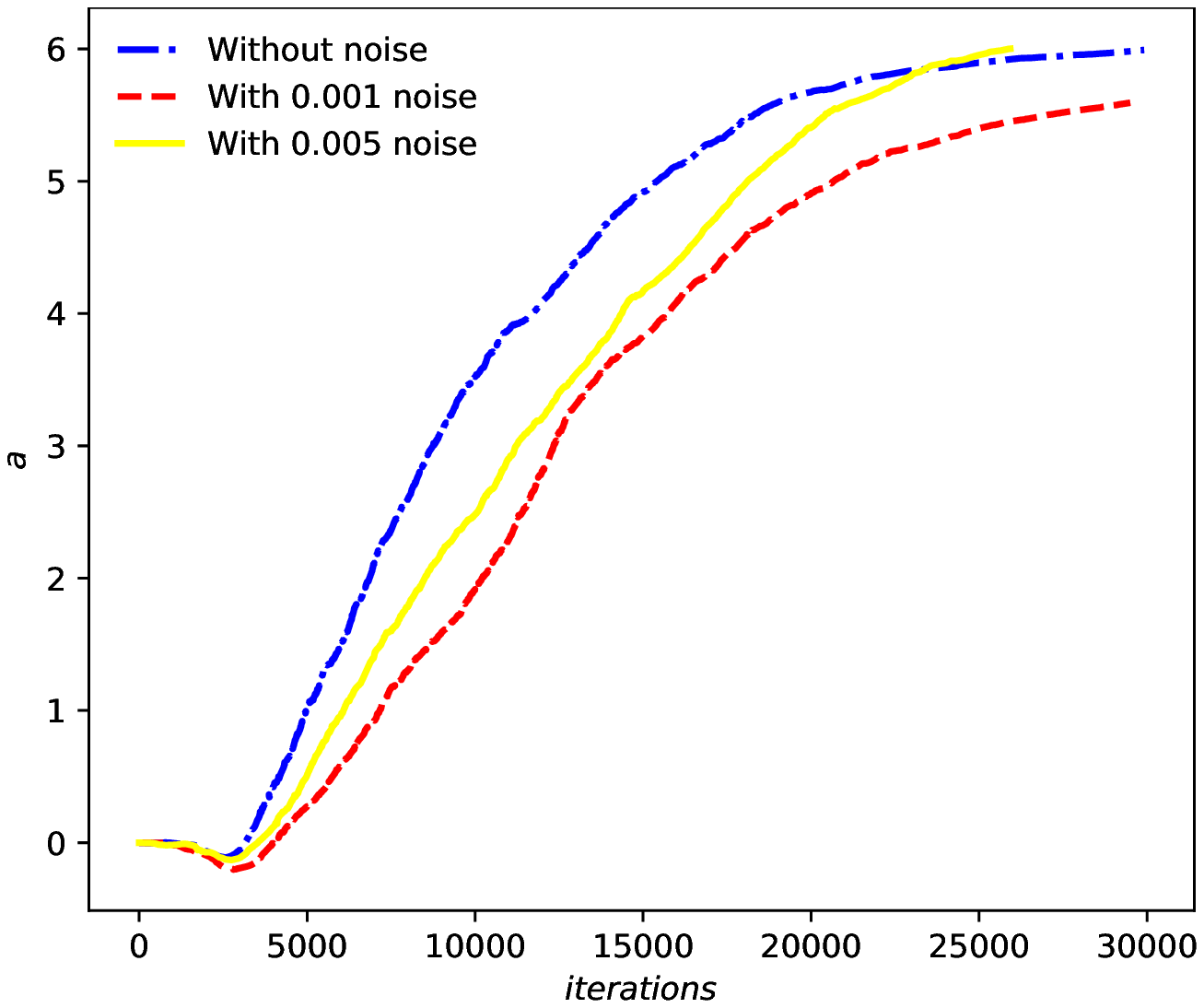}
\includegraphics[width=7cm,height=5cm]{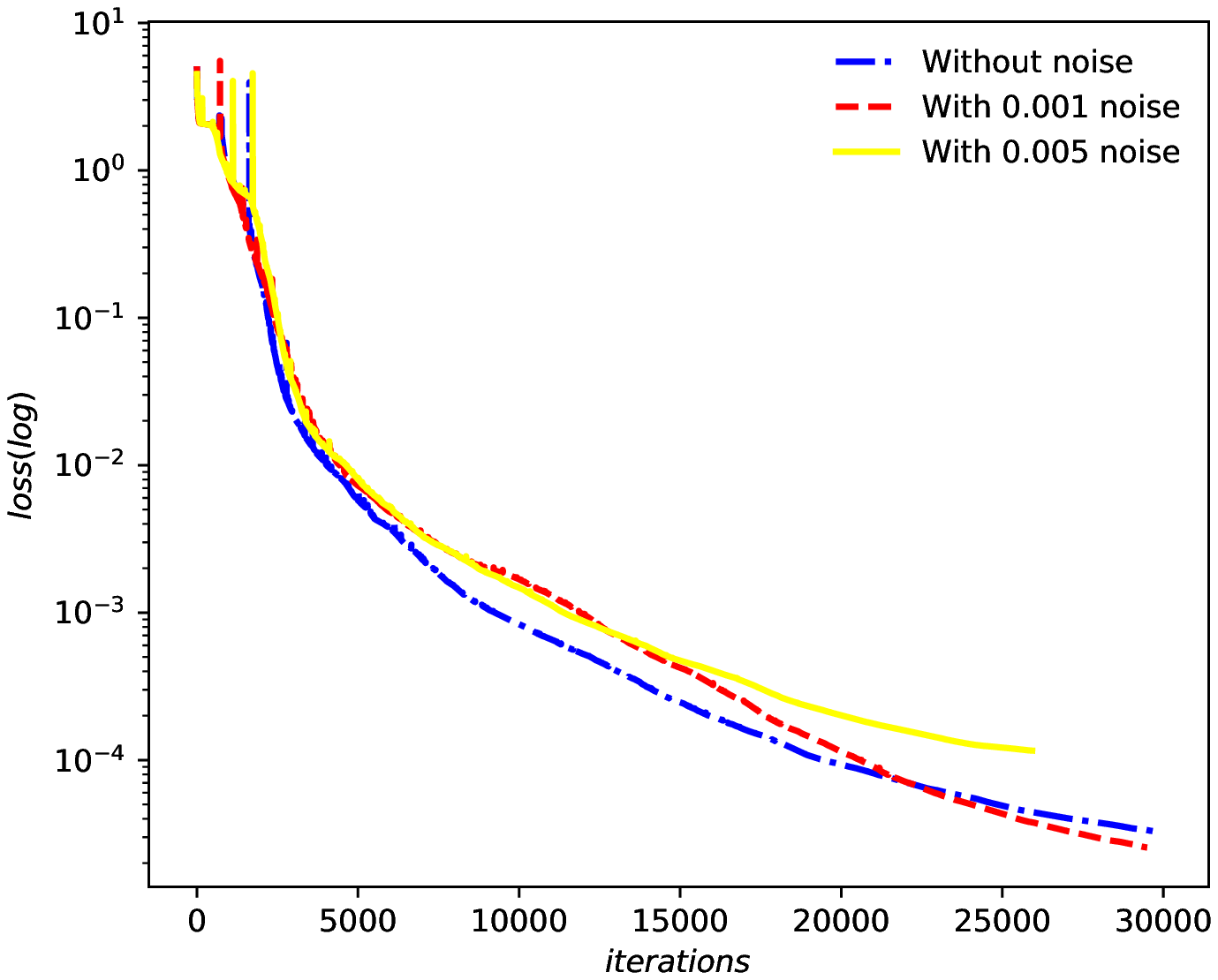}\\
$(c)$ \quad \quad \quad \quad \quad \quad \quad \quad \quad \quad \quad $(d)$
\caption{(Color online)  Parameter discovery of the nonlocal mKdV equation:(a) Iteration diagram of different internal configuration points;
(b) Iterative graph of different neurons in each layer with the same neural network depth;
(c) Iterative graph under different noise conditions;
(d) The variation of loss function with the different noise.}\label{9t}
\end{figure}

In Figures \ref{9t} (a) and  (b), we show the iterative changes of unknown parameters in the process of inverse problem under different internal configuration points and different number of neurons. When other conditions are determined, the more internal configuration points or the more neurons in each layer, the faster the convergence of unknown parameters. In Fig. \ref{9t} (c) and (d), the changes of unknown parameters and loss functions with iteration are analyzed when different noises are used in the inverse problem. The iteration of unknown parameters under different noises is described in Fig. \ref{9t} (c). The convergence without noise is faster than that with noise, but the final learning result is better when the noise is 0.5. Fig. \ref{9t} (d) describes the change of loss functions of different noises with the increase of iteration times. The results show that the convergence effect is the best when $0.05\%$ noise is added to the PINN. When $0.01\%$ noise is added to the network, the convergence effect is the worst. This shows that adding proper noise into the network is beneficial to improving the training results.

\begin{table}[H]
\caption{Parameter discovery through different neurons with different noises.}
\label{tableA}
\begin{tabular}{|l|ll|ll|ll|}
\hline
\multirow{2}{*}{\begin{tabular}[c]{@{}c@{}}Hidden layers\\ -Neurons \end{tabular}}           & \multicolumn{2}{c|}{Without noise}                         & \multicolumn{2}{c|}{With a 0.1\% noise}    &\multicolumn{2}{c|}{With a 0.5\% noise}                    \\ \cline{2-7}
\multicolumn{1}{|c|}{} & \multicolumn{1}{c}{a}         & \multicolumn{1}{c|}{error} & \multicolumn{1}{c}{a}         & \multicolumn{1}{c|}{error} & \multicolumn{1}{c}{a}         & \multicolumn{1}{c|}{error} \\ \hline
~~~~~~~~~~~~~~~9-40                     & \multicolumn{1}{l|}{5.385962} & 10.23397\%                 & \multicolumn{1}{l|}{5.941454} & 0.97576\%                  & \multicolumn{1}{l|}{5.86068}  & 2.32201\%                  \\ \hline
~~~~~~~~~~~~~~~9-60                     & \multicolumn{1}{l|}{5.817094} & 3.04843\%                  & \multicolumn{1}{l|}{5.822274} & 2.9621\%                   & \multicolumn{1}{l|}{5.6708}   & 5.48667\%                  \\ \hline
~~~~~~~~~~~~~~~9-80                     & \multicolumn{1}{l|}{5.991696} & 0.1384\%                   & \multicolumn{1}{l|}{5.593484} & 6.77526\%                  & \multicolumn{1}{l|}{6.001476} & 0.0246\%                   \\ \hline
\end{tabular}
\end{table}

\begin{table}[H]
\caption{Parameter discovery through different internal configuration points with different noises.}
\label{table2}
\begin{tabular}{|l|ll|ll|ll|}
\hline
\multirow{2}{*}{\begin{tabular}[c]{@{}c@{}}Internal points\\ $N_f$ \end{tabular}}  & \multicolumn{2}{c|}{Without noise}                                & \multicolumn{2}{c|}{With a 0.1\% noise}                           & \multicolumn{2}{c|}{With a 0.5\% noise}                           \\ \cline{2-7}
\multicolumn{1}{|c|}{} & \multicolumn{1}{c}{a}         & \multicolumn{1}{c|}{error} & \multicolumn{1}{c}{a}         & \multicolumn{1}{c|}{error} & \multicolumn{1}{c}{a}         & \multicolumn{1}{c|}{error}  \\ \hline
10000                                     & \multicolumn{1}{l|}{5.635974}   & 6.0671\%                        & \multicolumn{1}{l|}{5.866333}   & 2.22778\%                       & \multicolumn{1}{l|}{5.132474}   & 14.45876\%                      \\ \hline
30000                                     & \multicolumn{1}{l|}{5.702411}   & 4.95981\%                       & \multicolumn{1}{l|}{5.5218}     & 7.97\%                          & \multicolumn{1}{l|}{5.752123}   & 4.13129\%                       \\ \hline
50000                                     & \multicolumn{1}{l|}{5.991696}   & 0.1384\%                        & \multicolumn{1}{l|}{5.593484}   & 6.77526\%                       & \multicolumn{1}{l|}{6.001476}   & 0.0246\%                        \\ \hline
\end{tabular}
\end{table}

\section{Conclusion}
In this paper, we first give the infinite conservation laws of the nonlocal mKdV equation by using Riccti-type equation, which shows the integrability of the nonlocal mKdV equation. Then the nonlocal term is added to the PINN deep learning to reconstruct the zero boundary solutions of the nonlocal mKdV equation, including kink solution, complex soliton solution, bright-bright soliton  and bright-kink interaction solutions; For non-zero boundary, we use PINN to numerically simulate kink solution, dark soliton solution, anti-dark soliton solution and rational solution. At the same time, the error of each learning solution under $\mathbb{L}_2$ norm is given. The results show that the error between the exact solution and the exact solution generated by the PINN deep learning method is very small, which verifies the effectiveness and stability of the integrable deep learning method. For a given region, PINN can quickly and accurately learn the corresponding solution by using less data and network layers, which shows the power of integrable deep learning. Finally, the problem of data-driven parameter discovery is solved. The coefficients of nonlinear terms for the nonlinear mKdV equation are learned through PINN. In the process of inverse problem processing, we add noise to the network. Using the control variable method, when the training data and network depth are determined, we compare the parameter discovery of different neurons in each hidden layer under different noise conditions, and present the results in the form of a table. Under the condition of determining the network depth and neurons, we compare the learning results of parameters given different training data and noise. The results show that the more internal configuration points are given, the more accurate the learning results are. In addition, the discovery of parameters is also closely related to noise. The addition of different noises has a great influence on the learning results, which indicates that noise has a strong sensitivity.

\section*{Acknowledgments}
The project is supported by National Natural Science Foundation of China (No. 12175069 and No. 12235007) and Science and Technology Commission of Shanghai Municipality (No. 21JC1402500 and No. 22DZ2229014).

\end{document}